\documentclass[a4paper,11pt]{article}
\usepackage[utf8]{inputenc}
\usepackage[a4paper, total={6.5in, 8in}]{geometry}
\usepackage{graphicx}  
\usepackage{dcolumn}  
\usepackage{bm}        
\usepackage{amssymb}
\usepackage{hyperref}
\usepackage{url}
\usepackage{multirow}
\usepackage{amsmath}
\usepackage{cases}
\usepackage{placeins}
\usepackage{float}
\usepackage{subcaption}
\usepackage{color}
\usepackage{authblk}
\usepackage[table]{xcolor}
\usepackage[backend=biber,style=ieee,sorting=none]{biblatex}

 \hypersetup{
     colorlinks=true, 
     linktoc=all,     
     linkcolor=magenta,  
     citecolor= blue
}
\usepackage{tikz}
\usepackage[compat=1.1.0]{tikz-feynman} 

\newcommand{\GeV}{{\, \rm GeV}}
\newcommand{\TeV}{{\, \rm TeV}}
\newcommand{\eps}{\epsilon}

\newcommand{\gm}{\gamma^\mu}
\newcommand{\gmcon}{\gamma_\mu}

\newcommand{\mn}{\mu\nu}

\newcommand{\sigsubmn}{\sigma_{\mu\nu}}

\newcommand{\sw}{s_{\mathrm{w}}}
\newcommand{\cw}{c_{\mathrm{w}}}

\newcommand{\be}{\begin{equation}} 
\newcommand{\ee}{\end{equation}} 
\newcommand{\bea}{\begin{eqnarray}}
\newcommand{\eea}{\end{eqnarray}}
\newcommand{\bee}{\begin{eqnarray*}}
\newcommand{\eee}{\end{eqnarray*}}

\newcommand{\cO}{{\mathcal O}} 
\newcommand{\cL}{{\mathcal L}}  
 
\newcommand{\cH}{{\mathcal H}}
\newcommand{\cB}{{\mathcal B}}

\newcommand{\Mueg}{\mu\rightarrow e\gamma}
\newcommand{\Mueee}{\mu\rightarrow eee}
\newcommand{\Taueg}{\tau\rightarrow e\gamma}
\newcommand{\Taumug}{\tau\rightarrow \mu\gamma}
\newcommand{\CRmue}{\rm CR (\mu\rightarrow e)}

\newcommand{\Bsmue}{B_s\to \mu^+e^- }
\newcommand{\BK}{B^+\to K^+\mu^+e^-}
\newcommand{\BKs}{B\to K^*\mu^+e^-}
\newcommand{\BksZ}{B^0\to K^{*0}\mu^+ e^-}
\newcommand{\Bsphi}{B_s\to \phi\mu^-e^+}

\def\e{e}

\renewcommand{\bar}{\overline}

\newcommand{\BR}{\text{BR}}
\newcommand{\Br}{\text{Br}}

\newcommand{\nnu}{\nonumber\\}

\def\Cone{\boldsymbol{C_1}}
\def\Ctwo{\boldsymbol{C_2}}
\def\Cthree{\boldsymbol{C_3}}
\def\Cfour{\boldsymbol{C_4}}

\def\lQ1{C_{\ell q}^{(1)}}
\def\lq3{C_{\ell q}^{(3)}}
\def\qe{C_{qe}}
\def\phie{C_{\vp e}}
\def\Phil1{C_{\vp \ell}^{(1)}}
\def\phil3{C_{\vp \ell}^{(3)}}
\def\ld{C_{\ell d}}
\def\ed{C_{ed}}

\def\ledq{C_{\ell edq}}
\def\ledqp{C_{\ell edq}^{\prime}}
\def\mesonV{\mathcal V}
\def\mesonP{\mathcal P}

\newcommand{\vp}{\varphi}

\newcommand{\vpj }{\mbox{${\vp^\dag \,\raisebox{1.2mm}{\boldmath ${}^\leftrightarrow$}\hspace{-4mm} D_\mu\,\vp}$}}
\newcommand{\vpjt}{\mbox{${\vp^\dag \,\raisebox{2mm}{\boldmath ${}^\leftrightarrow$}\hspace{-4mm} D_\mu^{\,I}\,\vp}$}}

\title{SMEFT analysis of charged lepton flavor violating $B$-meson decays}

\author{Md Isha Ali\thanks{\href{mailto:isha.ali080@gmail.com}{isha.ali080@gmail.com}}, Utpal Chattopadhyay\thanks{\href{mailto:tpuc@iacs.res.in}{tpuc@iacs.res.in}}, N Rajeev\thanks{\href{mailto:spsrn2733@iacs.res.in}{spsrn2733@iacs.res.in}} and Joydeep Roy\thanks{\href{mailto:spsjr2729@iacs.res.in}{spsjr2729@iacs.res.in}}}

\affil{School of Physical Sciences, Indian Association for the Cultivation of Science, \\ 2A \& 2B, Raja SC Mallick Rd, Jadavpur, Kolkata - 700032, India.}

\date{}

\DeclareUnicodeCharacter{2212}{-}

\begin{document}

\maketitle

\begin{abstract}
Charged lepton flavor violation (cLFV) processes, potentially important for various Beyond the Standard Model Physics
scenarios are analyzed in the Standard Model Effective Field Theory (SMEFT) framework. We consider the most relevant 2 quark-2 lepton $(2q2\ell)$ operators for the leptonic and semi-leptonic LFV B-decay  (LFVBD) processes  $\Bsmue, \BK, \BksZ, {\rm and}~ \Bsphi$. We analyse the interplay among the Wilson coefficients responsible for these LFVBDs and other cLFV processes like $\CRmue$, $\ell_i \to \ell_j \gamma$, $\ell_i \to\ell_j\ell_k\ell_m$ and $Z \to \ell_i \ell_j$, to find the maximal possible LFV effects in $B$-meson decays. We probe the scale of new physics in relation to the 
constraints imposed by both classes of the LFV decays while considering  both the present bounds and future expectations. In view of proposed experiments at LHCb-II and Belle II to study charged LFV processes, we have also provided the upper limits on the indirect constraints on such LFVBDs. For the processes where $B$ meson is decaying to $\mu^{\pm}$ and $e^{\mp}$, we show that new physics can be constrained by an enhancement of 2-4 orders of magnitude on the current sensitivities of the BRs of $\BK, \BksZ {\rm and}~ B_s\to\phi\mu^{\pm}e^{\mp}$.
\end{abstract}



\section{Introduction}
\label{Sec:Intro}

The Standard Model (SM)~\cite{Peskin:1995ev} has been extraordinarily successful in elucidating the fundamental interactions between constituent particles.
It has made precise predictions that have been verified by experiments at accelerators such as LEP~\cite{Myers:1990sk}, Tevatron~\cite{Wilson:1977nk}, and the LHC~\cite{Evans:2008zzb}. The discovery of the Higgs boson at the LHC in 2012~\cite{ATLAS:2012yve} was the last missing piece of the SM puzzle, and it firmly establishes the SM as the appropriate theory for the energy range we have explored. However, SM is still far from becoming a comprehensive account of particle physics~\cite{Crivellin:2023zui}. There are several experimental facts and theoretical questions that cannot be addressed by staying within the SM framework. These include the gauge hierarchy problem, the mass of neutrinos, the lack of a particle dark matter (DM) candidate, and baryon asymmetry of the universe, among other significant problems that drive our investigation of physics Beyond the SM (BSM) scenarios.
Direct and indirect avenues exist for exploring the potential existence of New Physics (NP). Direct approaches involve detecting new particles through ongoing and upcoming collider experiments, while indirect probes rely on evidence gathered from various low-energy processes. Among the many potential BSM signals, lepton flavor violation (LFV) stands out as an intriguing and promising candidate for investigating NP scenarios. 

The SM assumption of the left-handed neutrinos to be massless renders a lepton family number  to be a conserved quantity, but the neutrino oscillation experiments confirmed that the lepton flavor conservation is not a  symmetry of nature~\cite{Super-Kamiokande:1998kpq,SNO:2002tuh}. In the minimum extension of SM where neutrinos have non-vanishing masses, charged lepton flavor violation (cLFV) is enabled via neutrino oscillation. But, it is heavily suppressed\footnote{By a factor proportional to ($\Delta m^2/m_W^2)^2 \sim 10^{-50}$, $\Delta m^2$ being the squared mass differences of the neutrino mass eigenstates.} by the Glashow–Iliopoulos–Maiani (GIM) mechanism, making them unobservable in the current experiments~\cite{Calibbi:2017uvl,Ardu:2022sbt}. We will briefly review the present experimental status of cLFV processes across various sectors, encompassing lepton decays, boson decays, and hadronic decays.

Before the discovery of neutrino oscillation, cLFV was searched in the decays of atmospheric muons without neutrinos~\cite{Hincks:1948vr}. cLFV in muons are well studied in the decays such as $\mu^+ \to e^+\, \gamma$, $\mu^+ \to e^+ e^- e^+$ and $\mu \to e$ conversion via various nuclei (Au, Al, Ti).
MEG~\cite{MEG:2016leq,MEGII:2018kmf} and PSI~\cite{SINDRUM:1987nra,SINDRUMII:2006dvw} experiments provide the current limit on $\mu^+ \to e^+\, \gamma$, with MEG setting an upper limit of $4.2 \times 10^{-13}$ at 90 \% confidence level (CL)~\cite{MEG:2016leq}. This bound is the strongest known in the LFV sector. PSI and upgraded MEG II aim to double the muon measurement rate~\cite{MEGII:2018kmf}. The SINDRUM experiment at PSI sets the limit for $\mu^+ \to e^+ e^- e^+$ at $<1.0 \times 10^{-12}$ at 90\% CL~\cite{SINDRUM:1987nra}. Mu3e at PSI targets an upper limit of $10^{-16}$ by 2030~\cite{Wauters:2021ghc}. Similarly, the best upper bound on neutrinoless $\mu \to e$ conversion using gold targets is $7 \times 10^{-13}$ by SINDRUM II at PSI~\cite{SINDRUMII:2006dvw}. Mu2e at Fermilab projects sensitivity to $\cO(10^{-17})$ using aluminum targets~\cite{Mu2e:2014fns}, while COMET at J-PARC aims for $\cO(10^{-15})$ and $\cO(10^{-17})$ sensitivities in phases 1 and 2 with the same target~\cite{COMET:2018auw}. DeeMe at J-PARC projects $\cO(10^{-13})$ sensitivity using Silicon Carbide (SiC) targets~\cite{Natori:2014yba}.
In the $\tau$-sector, constraints on $\tau$-cLFV decays are challenging due to lower production rates and shorter lifetime. Moreover, the constraints from $\tau$-cLFV are not as stringent as those for the case of muon. On the other hand, $\tau$-cLFV decays have the potential for neutrinoless semileptonic decays. B-factories, such as BaBar~\cite{BaBar:2001yhh} and Belle~\cite{Belle:2000cnh}, have set upper limits on branching fractions like $\tau^- \to \mu^- \gamma$ and $\tau^- \to e^- \gamma$~\cite{BaBar:2009hkt,Belle:2021ysv}. Belle-II aims for greater sensitivity~\cite{Belle-II:2018jsg}. In addition, $\tau^- \to \ell^- \ell^+ \ell^-$ decays are background-free and attractive for LHC experiments, with the strongest limit from LHCb~\cite{LHCb:2014kws}. Apart from the lepton sector, there are promising LFV searches in the bosonic sector involving Z~\cite{ATLAS:2021bdj,ATLAS:2022uhq} and Higgs bosons~\cite{ATLAS:2019pmk,CMS:2021rsq} that require high-energy colliders. $\mu$-cLFV and $\tau$-cLFV indirectly constrain the branching fraction of $Z/H \to \ell \ell'$ decays ($\ell^{(\prime)} \in e,\mu,\tau$)~\cite{Altmannshofer:2022fvz}. 

LFV in B-meson decays complements LFV searches in other sectors and these studies are strongly motivated 
by dedicated experiments, including LHCb and Belle II. Leptonic B-cLFV searches include decays like $B_{s}^0 \to e\mu$ and $B_{s}^0 \to \tau \ell$, where $\ell \in e,\mu$. The LHCb experiment has established the most stringent upper limit on the branching fraction at $\mathcal{B}(B_s^0 \to e^\pm \mu^\mp)< 6.3 \times 10^{-9}$~\cite{LHCb:2017hag} at 90\% CL. Similarly, searching for $B_{s}^0 \to \tau \ell$ channels presents experimental difficulties due to missing energy from $\tau$ decays. Nonetheless, LHCb has achieved the strongest constraints on $B_{s}^0 \to \tau \ell$ branching fraction to be $<4.2 \times 10^{-5}$~\cite{LHCb:2019ujz} at 95\% CL. Semileptonic B-cLFV searches include well-known decays like $B \to K^{(*)}\ell \ell'$, with $\ell \in e,\mu,\tau$. The LHCb collaboration reported an exclusion limit of $\mathcal{B}(B^+\to K^+\mu^-e^+)<7 \times 10^{-9}$~\cite{LHCb:2019bix} at 90\% CL, based on Run I data with an integrated luminosity of 3 $\rm fb^{-1}$. Similar searches were performed for decays $B^0\to K^{*0}(\to K^+ \pi^-)\mu^{\pm}e^{\mp}$ and $B_s\to \phi (\to K^+ K^-)\mu^{\pm}e^{\mp}$ using LHCb data up to 13 TeV, corresponding to a total integrated luminosity of 9 $\rm fb^{-1}$~\cite{LHCb:2022lrd}. The current limits set by LHCb on the branching ratios (BR) are $\mathcal{B}(B^0\to K^{*0} \mu^{-}e^{+}) < 6.8 \times 10^{-9}$~\cite{LHCb:2022lrd} and $\mathcal{B}(B_s\to \phi \mu^{\pm}e^{\mp}) < 10.1 \times 10^{-9}$~\cite{LHCb:2022lrd} at 90\% CL. The involvement of $\tau$-leptons in the final states of these transitions leads to less stringent constraints due to the challenges posed by missing energy during reconstruction. However, with upgrades planned for LHCb (I and II)~\cite{LHCb:2018roe,LHCb:2022ine} and the full dataset expected from Belle II~\cite{Belle-II:2018jsg} by the end of this decade, there is a possibility of improving the upper limits on these processes by up to an order of magnitude. In Table \ref{Tab:LFVlimits}, we have compiled the current and future prospects for all the LFV processes discussed above. In view of these present and predicted future measurements, in this work, we plan to perform an assessment of the maximal possible LFV effects in $B$-meson decays in a model-independent way.

In the model-dependent category, various popular models have been used to accommodate cLFV processes. These models include the Two Higgs Doublet Model (2HDM)\cite{Davidson:2016utf,Diaz:2000cm,Diaz:2002uk,Chang:1993kw,Paradisi:2006jp}, Supersymmetric (SUSY) extensions of the SM \cite{Barbieri:1995tw,Hisano:1995cp,Hisano:1996qq,Hisano:1998fj,Ellis:1999uq,Casas:2001sr,Calibbi:2006nq,Calibbi:2012gr,Hirsch:2012ti,Calibbi:2015kja,Evans:2018ewb,Hirao:2021pmh}, the Minimal Supersymmetric Standard Model (MSSM) \cite{Chattopadhyay:2019ycs}, the seesaw mechanism, which explains neutrino masses and mixing, leading to cLFV processes~\cite{Girrbach:2009uy,Masina:2002mv,Goto:2014vga,Vicente:2015cka,Bonilla:2016fqd,Calibbi:2017uvl} or flavor symmetry models ~\cite{Altmannshofer:2014cfa}. Since the last decade, anomalies in $B$ decays proceeding via $b\to s \ell_i\ell_j$ quark level transition have drawn significant attention due to their association with lepton flavor universality violation (LFUV) based on symmetry arguments as provided in  Ref.~\cite{Glashow:2014iga}. Various BSM models, which aim to explain LFUV\footnote{ Our analysis in particular hardly includes any LFUV studies.}, also introduce the possibility of LFV in B decays, have been studied extensively \cite{Alonso:2015sja,Greljo:2015mma,Boucenna:2015raa,Falkowski:2015zwa,Guadagnoli:2016erb,Feruglio:2016gvd}. Besides, such LFVBDs appear in LQ models~\cite{deMedeirosVarzielas:2015yxm,Sahoo:2015pzk,Duraisamy:2016gsd, Becirevic:2016oho,Crivellin:2017dsk,Sheng:2018qtp, Kumbhakar:2022szr}, extended gauge sectors \cite{Crivellin:2015mqa,Crivellin:2015era,Becirevic:2016zri,Fayyazuddin:2018zww} or SUSY model \cite{Sheng:2018ylg}. 

To date, searches at the LHC have not yielded any  direct evidence of a new particle near the electroweak scale. Strong arguments in support of BSM physics 
and these null results can become the motivation of considering an Effective Field Theory (EFT) approach \cite{Georgi:1993mps,Buchmuller:1985jz,kaplan:eft,Bechtle:2022tck} to estimate the level of unknown physics interactions. In contrast to  considering a BSM model that is associated with a top-down approach to  an EFT framework, one can adopt a bottom-up approach\cite{kaplan:eft,Bechtle:2022tck} and here, this refers to the model-independent investigation within the Standard Model Effective Field Theory (SMEFT)\cite{kaplan:eft,Bechtle:2022tck,Buchmuller:1985jz,Grzadkowski:2010es} where the energy scale $\Lambda$ of effective interactions can be above the reach of current experiments.  In SMEFT one considers higher-dimensional effective local operators out of SM fields only. The operators respect SM gauge invariance, and they are suppressed by appropriate powers of $\Lambda$. In regard to LFV processes, SMEFT is shown to be a useful tool for estimating 
any new physics effect at the scale $\Lambda$~\cite{Davidson:2018kud,Crivellin:2017rmk,Cirigliano:2017azj,Davidson:2017nrp,Davidson:2020ord,Davidson:2020hkf,Cirigliano:2021img,Kumar:2021yod}. For B-meson decays in particular, such model independent approaches have been implemented in a few works, a partial set of references are \cite{Calibbi:2015kma,Descotes-Genon:2023pen}.
  
\begin{table}[t!]
\centering
\resizebox{\textwidth}{!}{
\renewcommand{\arraystretch}{1.2}
\begin{tabular}{|l|cl|cl|}
\hline
Observables of cLFV modes. & \multicolumn{2}{c}{Present bounds} & \multicolumn{2}{|c|}{Expected future limits} \\
\hline
\hline
BR$(\mu\to e\gamma)$ &  $4.2\times10^{-13}$ & MEG(2016)~\cite{MEG:2016leq} & $6\times 10^{-14}$ & MEGII\cite{MEGII:2018kmf}\\
BR$(\mu\to eee)$ &   $1.0\times10^{-12}$ & SINDRUM(1988)~\cite{SINDRUM:1987nra} & $10^{-16}$ & Mu3e\cite{Blondel:2013ia} \\
\hline
CR$(\mu-e,{\rm Au})$ & $7.0\times10^{-13}$ & SINDRUMII(2006)~\cite{SINDRUMII:2006dvw}&  --\hspace{.7cm}  & -- \\
\multirow{2}{*}{CR$(\mu-e,{\rm Al})$} & - & - & $6\times10^{-17}$ & COMET/Mu2e\cite{Kuno:2013mha,Mu2e:2014fns} \\
 & - &          -              & $10^{-15}$ (Phase I) \& $10^{-17}$ (Phase II) & J-PARK\cite{COMET:2018auw} \\
\hline\hline
BR$(\tau\to e\gamma)$ & $3.3\times10^{-8}$ &  BaBar(2010)~\cite{BaBar:2009hkt} & $ 3\times 10^{-9\phantom{0}}$ & Belle-II\cite{Belle-II:2018jsg}\\
BR$(\tau\to eee)$ & $2.7\times10^{-8}$ & BaBar(2010)~\cite{Hayasaka:2010np} & $5\times 10^{-10}$ & Belle\cite{Belle:2007cio}\\
BR$(\tau\to e\mu\mu)$ & $2.7\times10^{-8}$ &  BaBar(2010)~\cite{Hayasaka:2010np} & $5\times 10^{-10}$ & Belle-II\cite{Belle-II:2018jsg}\\
\hline
BR$(\tau\to \mu\gamma)$ & $4.2\times10^{-8}$ &  Belle(2021)~\cite{Belle:2021ysv} & $ 10^{-9\phantom{0}}$ & Belle-II\cite{Belle-II:2018jsg}\\
BR$(\tau\to \mu\mu\mu)$ & $2.1\times10^{-8}$ &  BaBar(2010)~\cite{Hayasaka:2010np} & $4\times 10^{-10}$ & Belle-II\cite{Belle-II:2018jsg}\\
BR$(\tau\to \mu ee)$ & $1.8\times10^{-8}$ & BaBar(2010)~\cite{Hayasaka:2010np} & $3\times 10^{-10}$ & Belle-II\cite{Belle-II:2018jsg}\\
\hline
BR$(\tau\to \pi\mu)$ & $1.1\times10^{-7}$ & BaBar(2006)~\cite{BaBar:2006jhm} & $5\times 10^{-10}$ & Belle-II\cite{Belle-II:2018jsg}\\
BR$(\tau\to \rho\mu)$ & $1.2\times10^{-8}$ & BaBar(2011)~\cite{Belle:2011ogy} & $2\times 10^{-10}$ & Belle-II\cite{Belle-II:2018jsg}\\
\hline\hline
$\text{BR}(Z\to \mu e)$  &  $ 1.7\times 10^{-6}$ LEP (95\% CL) \cite{OPAL:1995grn} & \hspace{-.45cm}  $ 7.5\times 10^{-7}$ LHC (95\% CL)\cite{ATLAS:2014vur} & $10^{-8}$\,--\,$10^{-10}$ & {CEPC\cite{CEPCStudyGroup:2018ghi,CEPCStudyGroup:2018rmc}/FCC-ee\cite{Benedikt:2653673,Hernandez-Tome:2019lkb}} \\
$\text{BR}(Z\to \tau e)$ &   $ 9.8\times 10^{-6}$\quad\cite{OPAL:1995grn} &   $ 5.0\times 10^{-6}$\quad\cite{ATLAS:2014vur,ATLAS:2020zlz} & $10^{-9}$ & {CEPC\cite{CEPCStudyGroup:2018ghi,CEPCStudyGroup:2018rmc}/FCC-ee\cite{Benedikt:2653673,Hernandez-Tome:2019lkb}}\\
$\text{BR}(Z\to \tau \mu)$ &  $ 1.2\times 10^{-5}$\quad\cite{DELPHI:1996iox} &   $ 6.5\times 10^{-6}$\quad\cite{ATLAS:2014vur,ATLAS:2020zlz}  & $10^{-9}$ & {CEPC\cite{CEPCStudyGroup:2018ghi,CEPCStudyGroup:2018rmc}/FCC-ee\cite{Benedikt:2653673,Hernandez-Tome:2019lkb}} \\
\hline
\hline
BR$(B^+\to K^+\mu^-e^+)$ & $ 7.0  (9.5)\times 10^{-9}$ &  LHCb(2019)~\cite{LHCb:2019bix} & $-$ & -\\
BR$(B^+\to K^+\mu^+e^-)$ & $ 6.4 (8.8)\times 10^{-9}$ &  LHCb(2019)~\cite{LHCb:2019bix} & $-$ & -\\
BR$(B^0\to K^{*0}\mu^+e^-)$ & $ 5.7  (6.9)\times 10^{-9}$ & LHCb(2022)~\cite{LHCb:2022lrd} & $-$ & -\\
BR$(B^0\to K^{*0}\mu^-e^+)$ & $ 6.8  (7.9)\times 10^{-9}$ & LHCb(2022)~\cite{LHCb:2022lrd} & $-$ & -\\
BR$(B^0\to K^{*0}\mu^{\pm}e^{\mp})$ & $ 10.1  (11.7)\times 10^{-9}$ & LHCb(2022)~\cite{LHCb:2022lrd} & $-$ & -\\
BR$(B_s^0\to \phi\mu^{\pm}e^{\mp})$ & $ 16  (19.8)\times 10^{-9}$ & LHCb(2022)~\cite{LHCb:2022lrd} & $-$ & -\\
\hline
BR$(B^+\to K^+\mu^-\tau^+)$ & $ 0.59\times 10^{-5}$ &  Belle(2022)~\cite{Belle:2022pcr} & $-$ & -\\
BR$(B^+\to K^{+}\mu^+\tau^-)$ & $ 2.45  \times 10^{-5}$ & Belle(2022)~\cite{Belle:2022pcr} & $3.3 \times 10^{-6}$ & Belle-II~\cite{Belle-II:2018jsg}\\
BR$(B^+\to K^{+}\tau^{\pm}e^{\mp})$ & $ 1.52  \times 10^{-5}$ & Belle(2022)~\cite{Belle:2022pcr} & $2.1 \times 10^{-6}$ & Belle-II~\cite{Belle-II:2018jsg}\\
BR$(B^0\to K^{*0}\tau^+\mu^-)$ & $1.0  (1.2)\times 10^{-5}$ & LHCb(2022)~\cite{LHCb:2022wrs} & $-$ & -\\
BR$(B^0\to K^{*0}\tau^-\mu^+)$ & $ 8.2  (9.8)\times 10^{-6}$ & LHCb(2022)~\cite{LHCb:2022wrs} & $-$ & -\\
\hline
BR$(B_{s}^0\to \mu^{\mp}e^{\pm})$ & $5.4 (6.3)\times 10^{-9}$ &  LHCb(2018)~\cite{LHCb:2017hag} & $3 \times 10^{-10}$ & LHCb-II~\cite{LHCb:2018roe}\\
BR$(B^0\to \mu^{\pm}e^{\mp})$ & $1.0(1.3)\times 10^{-9}$ & LHCb(2018)~\cite{LHCb:2017hag} & $-$ & $-$\\
BR$(B_s^0\to \tau^{\pm}e^{\mp})$ & $ 7.3\times 10^{-4} (95\%)$ &  LHCb(2019)~\cite{LHCb:2019ujz} & $-$ & $-$\\
BR$(B^0\to \tau^{\pm}e^{\mp})$ & $ 2.1\times 10^{-5} (95\%)$ &  LHCb(2019)~\cite{LHCb:2019ujz} & $-$ & $-$\\
BR$(B_s^0\to \tau^{\pm}\mu^{\mp})$ & $ 4.2\times 10^{-5} (95\%)$ &  LHCb(2019)~\cite{LHCb:2019ujz} & $-$ & $-$\\
BR$(B^0\to \tau^{\pm}\mu^{\mp})$ & $ 1.4\times 10^{-5} (95\%)$ &  LHCb(2019)~\cite{LHCb:2019ujz} & $1.3 \times 10^{-6}$ & Belle-II~\cite{Belle-II:2018jsg}\\
\hline\hline
\end{tabular}}
\caption{Present upper bounds (with $90\%\,\text{CL}$, unless otherwise specified), and future expected sensitivities of branching ratios for the set of low-energy cLFV transitions relevant for our analysis. For LFVBDs, the numbers within the parenthesis represent the results obtained with $95\%\,\text{CL}$. \label{Tab:LFVlimits}}
\end{table}

In this work, we study cLFV decays with more  emphasis on the leptonic and semileptonic B-decays in the SMEFT formalism consisting of dimension-6 operators. We also check for indirect constraints on the important Wilson coefficients (WCs) coming from some LFV processes other than those involving LFVBDs. The above particularly include limits from decays like $\CRmue$, $\ell_i \to \ell_j \gamma$, $\ell_i \to\ell_j\ell_k\ell_m$ and $Z \to \ell_i \ell_j$. We will henceforth collectively refer to these as ``other LFV processes". We estimate the effects on the WCs by including both current data  as well as future expectations from  measurements related to the other LFV processes, along with the same from LFVBDs. Thus, we  will also probe the interplay of different LFV bounds between the two sets of decays on the WCs considering the  prospective improved limits. We discuss scenarios constructed from different SMEFT coefficients by considering one operator 
at a time or turning on  
two operators to have non-vanishing WCs at the scale $\Lambda$ simultaneously. All other operators vanish at the same scale. 
As mentioned earlier, at the juncture of not receiving any NP result from the LHC, it is important to rely on indirect constraints from  LFVBDs, as well as the same from other LFV decays, and thereby limit the SMEFT operators. We are not aware of a comprehensive and updated analysis in this regard in relation to the above. This additionally motivates us to include the effects of considering at least two-order more stringent BRs for LFVBDs and explore these in relation to the possible future bounds of the other LFV processes.

The paper is organized as follows: in Sec.~\ref{Sec:SMEFT approach}, we give a general description of the SMEFT approach to cLFV, with more emphasis given to leptonic and B cLFV decays. In Sec.~\ref{Sec:Low-energy Eff. Hamiltonian}, we write down the most general low-energy effective Hamiltonian relevant for cLFV B decays and the effective Lagrangian for other cLFV decays, including $\Mueee$ and $\CRmue$. We also enumerate all the relevant SMEFT operators for current analysis and briefly discuss their correlation through renormalization group evolutions (RGE).  In Sec.~\ref{Sec:Results} we discuss our results by constructing various NP scenarios for different SMEFT operators in 1D and 2D analyses. Finally, we conclude in Sec.~\ref{Sec:Conclusion}.


\section{SMEFT approach to lepton flavor violation}
\label{Sec:SMEFT approach}
 
SMEFT describes new physics effects via higher dimensional operators, with mass dimension greater than 4 and consisting of SM fields at an energy scale $\Lambda$ that is above the reach of current experiments. 
The operators are suppressed by appropriate powers of $\Lambda~(>> m_W)$ and corresponding Wilson coefficients parameterise the low-energy behaviour of such high-energy theory through the running of RGEs of masses and coupling parameters of the theory. The SMEFT Lagrangian is thus given by \cite{Bechtle:2022tck,Grzadkowski:2010es}
\be \label{eq:SMEFT Lagrng} 
\cL_{\rm SMEFT} = \cL_{\rm SM} + \frac{1}{\Lambda} C^{(5)} \cO^{(5)} + \frac{1}{\Lambda^2} \sum_n C_n^{(6)} \cO_n^{(6)} + \cO (\frac{1}{\Lambda^3}) + ...
\ee
where $\cL_{\rm SM}$ is the usual renormalizable SM Lagrangian, $\cO^{(5)}$ represents the gauge-invariant mass dimension-5 operators, known as neutrino mass generating Weinberg operator, $C^{(5)}$ is the corresponding WCs. Similarly, $\cO_n^{(6)}$ and $C_n^{(6)}$ represent mass dimension-6 operators and corresponding WCs respectively. 
In this work, we will adopt
the conventions of the Warsaw basis \cite{Grzadkowski:2010es} and will 
not consider any more terms with suppression level greater than
$1/\Lambda^2$.

It is known that flavor-changing neutral current (FCNC) processes can be substantially large in many BSM scenarios, whereas they are heavily suppressed in the SM by small Cabibbo-Kobayashi-Maskawa (CKM)  matrix elements, loop effects etc. Probing NP models with FCNC effects can thus be quite useful. Therefore studying flavor observables in SMEFT approach which can provide sensitive NP contributions have been quite popular. For example, SMEFT has been used to study general LFV processes \cite{Crivellin:2013hpa,Pruna:2014asa,Crivellin:2017rmk}, $B$-meson LFV decays \cite{Alonso:2014csa, Aebischer:2015fzz, Crivellin:2015era, Becirevic:2016zri, Descotes-Genon:2023pen}, Z boson LFV decays \cite{Calibbi:2021pyh}, Higgs boson LFV decays \cite{Cullen:2020zof} or Quarkonium LFV decays \cite{Calibbi:2022ddo}. Although these references are hardly to be an exhaustive list of such works, it has been shown in all of them that most dominant contributions to LFV processes come from dimension-6 operators and furthermore the operators contributing to semi-leptonic processes are either of the Higgs-fermion or four-fermion type. With our focus on LFV $B$-meson decays only, we'll not consider the former type in our analysis. A complete list of dimension-six operators is provided in Table~\ref{tab:BtoS operators1} with the boldfaced ones contributing directly to LFVBDs. 

\begin{table}[t!] 
\centering
\renewcommand{\arraystretch}{1.3}
\setlength{\tabcolsep}{15pt}
\begin{tabular}{|cc|cc|} 
\hline
\multicolumn{2}{|c|}{4-lepton operators} & \multicolumn{2}{c|}{2-lepton-2-quark operators} \\
\hline
$\cO_{\ell\ell}$ & $(\bar L \gamma_\mu L)(\bar L \gamma^\mu L)$ & \bm{$\cO_{\ell q}^{(1)}$} & \bm{$(\bar L \gamma_\mu L)(\bar Q \gamma^\mu Q)$} \\
$\cO_{ee}$ & $(\bar E \gamma_\mu E)(\bar E \gamma^\mu E)$ & \bm{$\cO_{\ell q}^{(3)}$} & \bm{$(\bar L \gamma_\mu \tau^I L)(\bar Q \gamma^\mu \tau^I Q)$} \\
$\cO_{\ell e}$ & $(\bar L \gamma_\mu L)(\bar E \gamma^\mu E)$ & \bm{$\cO_{qe}$} & \bm{$(\bar Q \gamma_\mu Q)(\bar E \gamma^\mu E)$}  \\
\cline{1-2}
\multicolumn{2}{|c|}{Lepton-Higgs operators} & \bm{$\cO_{\ell d}$} & \bm{$(\bar L \gamma_\mu L)(\bar D \gamma^\mu D)$} \\
\cline{1-2}
$\cO_{\vp \ell }^{(1)}$ & $i(\vpj)(\bar L\gamma^\mu L)$ & \bm{$\cO_{ed}$} & \bm{$(\bar E \gamma_\mu E)(\bar D\gamma^\mu D)$} \\
$\cO_{\vp \ell}^{(3)}$ & $i(\vpjt)(\bar L \tau^I \gamma^\mu L)$ & \bm{$\cO_{\ell edq}$} & \bm{$(\bar L^a E)(\bar D Q^a)$} \\
$\cO_{\vp e}$ & $i(\vpj)(\bar E \gamma^\mu E)$ & $\cO_{\ell equ}^{(1)}$ & $(\bar L^a E) \eps_{ab} (\bar Q^b U)$ \\
$\cO_{e\vp }$ & $(\bar L E \Phi)(\Phi^\dag\Phi)$ & $\cO_{\ell equ}^{(3)}$ & $(\bar L^a \sigma_{\mu\nu} E) \eps_{ab} (\bar Q^b \sigma^{\mu\nu} U)$\\
\cline{1-2}
\multicolumn{2}{|c|}{Dipole operators} & $\cO_{\ell u}$ & $(\bar L \gamma_\mu L)(\bar U \gamma^\mu U)$  \\
\cline{1-2}
$\cO_{eW}$ & $(\bar L \sigma^{\mu\nu} E) \tau^I \Phi W_{\mu\nu}^I$ & $\cO_{eu}$ & $(\bar E \gamma_\mu E)(\bar U \gamma^\mu U)$  \\
$\cO_{eB}$ & $(\bar L \sigma^{\mu\nu} E) \Phi B_{\mu\nu}$ & & \\
\hline
\end{tabular}
\caption{A comprehensive list of dimension-6 operators that remain invariant under the SM gauge group and contribute to LFV observables. Those boldfaced ones are mainly responsible for generating LFVBDs at the tree level.
In these expressions, $Q$ and $L$ represent left-handed quark and lepton $SU(2)$ doublets respectively with indices $a,b=1,2$. 
$U,\,D$ and $E$ denote right-handed up, down quark and lepton singlets, with $\Phi$ representing the Higgs doublet 
(and $\vpj\equiv \Phi^\dag (D_\mu \Phi)-(D_\mu \Phi)^\dag \Phi$).
$B_{\mu\nu}$ and $W^I_{\mu\nu}$ stand for the $U(1)$ and $SU(2)$ field strengths, respectively, while $\tau^I$ with $I=1,2,3$ denotes the Pauli matrices.
For brevity flavor indices are not explicitly shown in this list.
\label{tab:BtoS operators1}}
\end{table}

In general studies of low-energy observables in SMEFT, there are three energy scales. A high energy scale $\Lambda$, an intermediate energy scale of electroweak symmetry-breaking $m_W$ and a low-energy scale like $\sim m_b$ or $m_{\tau / \mu}$. Therefore for probing the level of contributions of the higher dimensional operators for LFV studies that may be consistent with experimental constraints, a general method of ``match and run" of RGEs is described below. At the first step, the SMEFT 1-loop RGEs \cite{Jenkins:2013zja, Jenkins:2013wua, Alonso:2013hga} of relevant WCs would be initialized at the scale $\Lambda (\sim \TeV)$ and run down to the electroweak scale $\sim m_{Z,W}$. The WCs under study are given a non-vanishing value like unity while other WCs are set to zero at the scale $\Lambda$ and RGE evolution is completed till the electroweak scale $m_W$. Of course, the WCs are hardly expected to remain at the their initial values including also the ones that were vanishing at the higher scale.
 This level of evolutions are adequate for the LFV decays of the $Z$ or Higgs bosons but not enough for processes referring to energies below the electroweak scale. Further down the scale, in the second step, as in a top-down approach of EFT, the heavy particles of the theory ($W^{\pm}$, $Z$, the Higgs boson and the top quark) are integrated out and the operators invariant under the QCD$\times$QED gauge groups and consisting of fields of light charged fermions $(u, d, c, s, b, e, \mu, \tau)$, neutral fermions $(\nu_e, \nu_{\mu}, \nu_{\tau})$ and gauge bosons 
$(F_{\mu\nu}, G_{\mu\nu}^a)$ describe the effective interactions. These operators are known as Low-Energy Effective Field Theory (LEFT) operators that contribute to the total LEFT Lagrangian containing dimension three and higher dimensional $(d>4)$ operators. The most relevant LEFT operators for our purpose are dimension six operators containing four fermions with at least one spinor combination consisting of two different charged lepton flavours \cite{Grinstein:1988me}. Schematically, these operators take the form 
\be \label{eq:LEFT oprtr}
\cO_{SAB} = (\bar \ell_i \Gamma_S P_A \ell_j)(\bar f_{\alpha} \Gamma_S P_B f_{\beta})
\ee
where $\ell_{i,j}$ are the lepton pairs, $f_{\alpha,\beta}$ are the fermion pairs, $P_{A,B}$ are the left and right projection operators and $\Gamma_S = \bf{1}, \gmcon ~\rm{and}~ \sigsubmn$ for scalar, vector and tensor respectively. Following the ``match and run" procedure, 1-loop
RGEs of these LEFT operators are matched at tree level to the SMEFT operators, the details of which can be found in Refs.\cite{Jenkins:2017jig, Jenkins:2017dyc}. In the third and final step of the process, the  running of these LEFT operators to the low-energy scale of $m_\tau, m_\mu$ or $m_b$ is performed to evaluate the desired experimental observables. We implement all these procedures in our numerical analysis with the help of the {\it wilson}~\cite{Aebischer:2018bkb} and {\it flavio}~\cite{Straub:2018kue} packages.  


\section{Low energy effective Hamiltonian and branching ratios for LFV decays}
\label{Sec:Low-energy Eff. Hamiltonian}

Keeping the prime focus on the lepton flavor violating $B$-meson decays, in this section, we shall discuss the effective Hamiltonian or Lagrangians for LFV processes such as $b\to s\ell_i\ell_j$, $\CRmue$ and $\ell_i \to \ell_j\ell_k\bar\ell_m$ that are relevant to our analysis. We shall also provide the expression for branching ratios of corresponding processes and that will be followed by the classification of higher-dimensional operators responsible for such processes. Being a relatively less important process for our analysis, the details of ZLFV is given in the Appendix~\ref{Appendix: ZLFV decays}.


\subsection{Lepton flavor violating $b\to s\ell_i\ell_j$ decays}
\label{Subsec: LFVB decays}

\begin{figure}[H]
\centering
\begin{subfigure}{0.4\textwidth}
\begin{tikzpicture}
\begin{feynman}
\vertex (a1) {\(\overline b\)};
\vertex[right=1.5cm of a1] (a2);
\vertex[right=1.5cm of a2] (a3);
\vertex[right=1.5cm of a3] (a4) {\(\overline s\)};
\vertex[above=3em of a1] (b1) {\(u\)};
\vertex[above=3em of a4] (b2) {\(u\)};
\vertex at ($(a2)!1.1!(a3)!1.1cm!60:(a3)$) (d);
\vertex[below=1.5em of a4] (c1) {\(\ell^+\)};
\vertex[below=3em of c1] (c3) {\(\ell^-\)};
\vertex[below right=2.5em of d] (c2);
\diagram* {
(a4) -- [fermion] (a3) -- [charged boson,edge label'=\(W^{+}\)] (a2) -- [fermion] (a1),
(b1) -- [fermion] (b2),
(a3) -- [fermion, quarter left] (d) -- [fermion, quarter left, edge label={\(\bar{u}\),\(\bar{c}\),\(\bar{t}\)}] (a2),
(d) -- [boson, edge label'={\(\gamma\),\(Z^{0}\)}] (c2),
(c1) -- [fermion] (c2),
(c2) -- [fermion] (c3),
};
\draw [decoration={brace}, decorate] (a1.south west) -- (b1.north west)
node [pos=0.5, left] {\(B^{+}\)};
\draw [decoration={brace}, decorate] (b2.north east) -- (a4.south east)
node [pos=0.5, right] {\(K^{+}\)};
\end{feynman}
\end{tikzpicture}
\caption{}
\label{Fig:BplusToKplusPingu}
\end{subfigure}
\hfill
\begin{subfigure}{0.4\textwidth}
\begin{tikzpicture}
\begin{feynman}
\vertex (a1) {\(\overline b\)};
\vertex[right=1.5cm of a1] (a2);
\vertex[right=2cm of a2] (a3);
\vertex[right=1.5cm of a3] (a4) {\(\overline s\)};
\vertex[above=3em of a1] (b1) {\(u\)};
\vertex[above=3em of a4] (b2) {\(u\)};
\vertex at ($(a2)!1.1!(a3)!1.1cm!60:(a3)$) (d);
\vertex[below=3.5em of a4] (c1){\(\ell^-\)};
\vertex[below=2em of c1] (c3) {\(\ell^+\)};
\vertex[below=2.0em of a3] (d1) ;
\vertex[below=2.0em of a2] (d2) ;
\diagram* {
(a4) -- [fermion] (a3) -- [fermion, edge label'={\(\bar{u}\),\(\bar{c}\),\(\bar{t}\)}] (a2) -- [fermion] (a1),
(b1) -- [fermion] (b2),
(a3) -- [charged boson,edge label=\(W^{-}\)] (d1),
(d2) -- [charged boson,edge label =\(W^{+}\)] (a2),
(d2) -- [fermion, edge label=\(\nu\)](d1),
(d1) -- [fermion] (c1),
(c3) -- [fermion] (d2),
};
\draw [decoration={brace}, decorate] (a1.south west) -- (b1.north west)
node [pos=0.5, left] {\(B^{+}\)};
\draw [decoration={brace}, decorate] (b2.north east) -- (a4.south east)
node [pos=0.5, right] {\(K^{+}\)};
\end{feynman}
\end{tikzpicture}
\caption{}
\label{Fig:BplusToKplusBox}
\end{subfigure}
\label{Fig:Feynman B decays}


\begin{subfigure}{0.5\textwidth}
\begin{tikzpicture}
\begin{feynman}
\vertex (a1) {\(\overline b\)};
\vertex[blob,shape=circle,minimum height=1.0cm,minimum width=0.5cm] at ($(a1)+ (3.0cm, 0.0cm)$)(a2){};
\vertex[right=3.0cm of a2] (a3) {\(\overline s\)};
\vertex[above=3em of a1] (b1) {\(u\)};
\vertex[above=3em of a3] (b2) {\(u\)};
\vertex[below=1.5em of a3] (c1) {\(\ell^{+}\)};
\vertex[below=3em of c1] (c3) {\(\ell^{\prime{-}}\)};
\diagram* {
(a3) -- [fermion] (a2), 
(a2) -- [fermion] (a1),
(b1) -- [fermion] (b2),
(c1) -- [fermion] (a2),
(a2) -- [fermion] (c3),
};
\draw [decoration={brace}, decorate] (a1.south west) -- (b1.north west)
node [pos=0.5, left] {\(B^{+}\)};
\draw [decoration={brace}, decorate] (b2.north east) -- (a3.south east)
node [pos=0.5, right] {\(K^{+}\)};
\end{feynman}
\end{tikzpicture}
\caption{}
\end{subfigure}
\caption{Figures (a) and (b) represent the Feynman diagrams for the SM processes of $B^+\rightarrow K^+\ell^+\ell^-$, while figure (c) represents the Feynman diagram for the process of $B^+\rightarrow K^+\ell^+\ell^{\prime -}$. The blob in figure (c) represents the effective vertex for the LFVBD processes produced by the higher-dimensional operators under consideration.}
\label{Fig:BplusToKplus}
\end{figure}
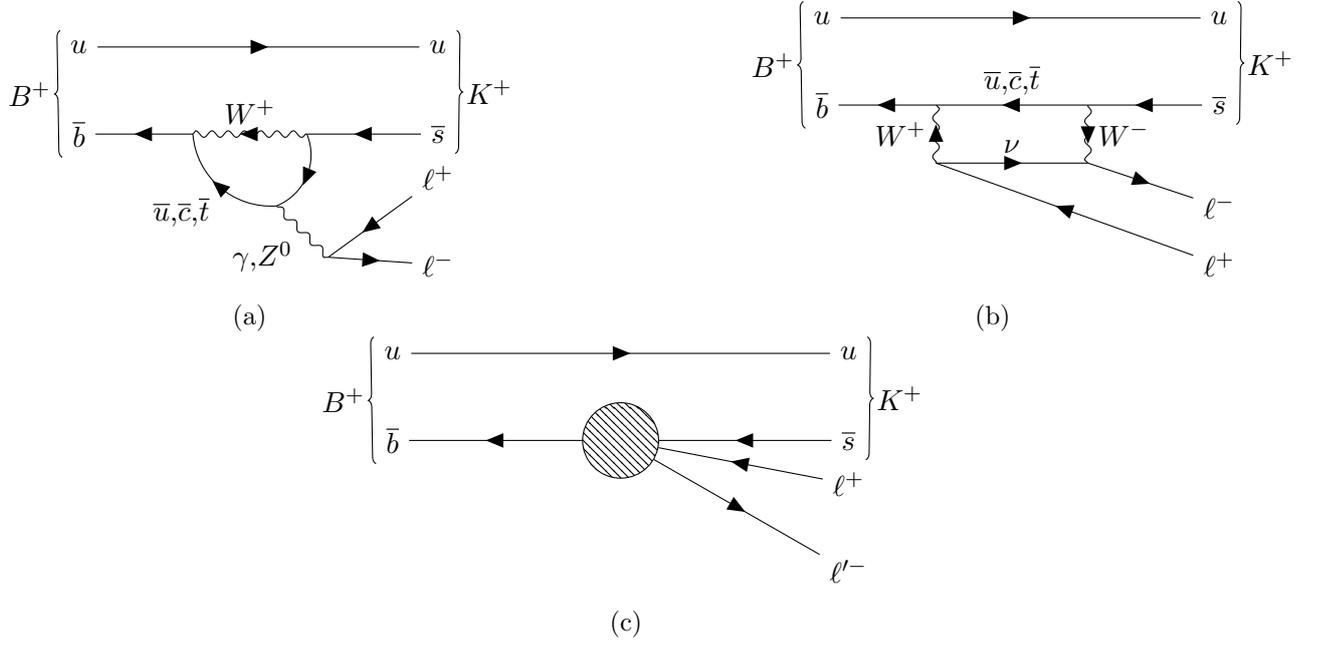

The most general effective Hamiltonian for the weak decay process of a bottom $(b)$ quark to strange $(s)$ quark transition along with two leptons $(\ell_{i,j})$, $b\to s\ell_i\ell_j$, in terms of low energy dim-6 operators $(\cO_n)$ and corresponding Wilson coefficients $(C_n)$ is given by \cite{Aebischer:2015fzz, Buchalla:1995vs} 
\be 
\label{eq: Eff Hamiltn}
\cH_{\rm eff}^{\Delta B=1} = - \dfrac{ 4 G_F }{\sqrt 2}
V_{tb}V_{ts}^* \bigg[\sum\limits_{n=7}^{\rm 10}{C_n (\mu) \cO_n (\mu) + C_n^\prime (\mu) \cO_n^\prime (\mu)} \bigg] \,,
\ee
where $G_F$ is the Fermi coupling constant, characterizing the strength of weak interactions, and $V_{tb}, V_{ts}^*$ are the CKM matrix elements. Both $\cO_n^{(\prime)}$ and $\cO_n^{(\prime)}$ are the functions of the renormalizable energy scale $\mu$, which for our low energy processes would be taken as the mass of the $b$ quark $(m_b)$. The primed operators are obtained by flipping the chirality, and they are usually highly suppressed compared to their unprimed counterparts in the SM. In Eq.~\ref{eq: Eff Hamiltn}, $n=7,8$ represent the photon and gluon ``Magnetic-Penguin" operators 
whereas, $n=9,10$ refer to the ``Semi-leptonic" operators. They are defined as \cite{Aebischer:2015fzz, Buchalla:1995vs}
\begin{align}
  \cO_7 & = 
    \frac{e}{16\pi^2} \overline m_b [\bar{s}_{Li}\sigma^{\mu\nu} b_{Ri}] F_{\mu\nu},\hspace{2 cm}
    \cO_8 = 
    \frac{g_s}{16\pi^2} \overline m_b [\bar{s}_{Li} \sigma^{\mu\nu} T^a_{ij}b_{Ri}] G^a_{\mn}, \label{eq:Mag-penguin Opr} \\
  \cO_9 & = 
    \frac{e^2}{16\pi^2} [\bar{s}_L \gamma_\mu b_L][\bar{\ell_i} \gamma^\mu \ell_j], \hspace{3 cm}
  \cO_{10}  = 
    \frac{e^2}{16\pi^2} [\bar{s}_L \gamma_\mu b_L][\bar{\ell_i} \gamma^\mu \gamma_5 \ell_j], \label{eq:Semi-leptonic Opr} 
\end{align}
where $\ell_{i,j}=e,\mu,\tau$ respectively. $F_{\mn}$ and  $G^a_{\mn}$ are the photon and gluon field strength tensor respectively and $R,L$ denote
the right and left-handed projection operators, $P_{R,L}=(1\pm\gamma_5)/2$. In Eq.~\ref{eq:Mag-penguin Opr}, $g_s$ is the strong coupling constant, $\overline m_b$ denotes the running of $b$ quark mass in the Minimal Subtraction $\overline {\text{(MS)}}$ scheme and $T^a_{ij}$ represents the color charges. $\cO_7, \cO_8, \cO_9$ and $\cO_{10}$ also have their chiral counterparts whose explicit structures are given by,

\begin{align}
  \cO_7^{\prime} & = 
    \frac{e}{16\pi^2} \overline m_b [\bar{s}_{Ri}\sigma^{\mu\nu} b_{Li}] F_{\mu\nu},\hspace{2 cm} \cO_8^{\prime} = 
    \frac{g_s}{16\pi^2} \overline m_b [\bar{s}_{Ri} \sigma^{\mu\nu} T^a_{ij}b_{Li}] G^a_{\mn},\label{eq:Mag-penguin-prm Opr} \\
  \cO_9^{\prime} & = 
   \frac{e^2}{16\pi^2} [\bar{s}_R \gamma_\mu b_R][\bar{\ell_i} \gamma^\mu \ell_j], \hspace{3 cm} ~
  \cO_{10}^{\prime}  = 
     \frac{e^2}{16\pi^2} [\bar{s}_R \gamma_\mu b_R][\bar{\ell_i} \gamma^\mu \gamma_5 \ell_j]. 
\end{align}  
In addition to these operators, two scalar, two pseudo-scalar and two tensor NP operators can potentially contribute to $B$-meson decays ~\cite{Aebischer:2015fzz, Bobeth:2007dw}, \footnote{ It is shown in \cite{Alonso:2014csa} that the contribution of tensor operators to such decays can be neglected because of not satisfying the of their non-invariant character under $SU(2)_L\times U(1)_Y$ 
symmetry.}
\begin{align}
  \cO_S & =  [\bar{s}_L b_R] [\bar{\ell_i} \ell_j],\hspace{2 cm}
  \cO_S^{\prime}  =  [\bar{s}_R b_L] [\bar{\ell_i} \ell_j], \\
  \cO_P & =  [\bar{s}_L b_R] [\bar{\ell_i} \gamma_5 \ell_j],\hspace{2 cm}
  \cO_P^{\prime}  =  [\bar{s}_R b_L] [\bar{\ell_i} \gamma_5 \ell_j],
\label{eq:nonSM:scalars} \\
  \cO_T   & =  [\bar{s} \sigma_{\mu\nu} b][\bar{\ell_i} \sigma^{\mu\nu} \ell_j], \hspace{1 cm}
  \cO_{T5}  =  [\bar{s} \sigma_{\mu\nu} b][\bar{\ell_i} \sigma^{\mu\nu} \gamma_5 \ell_j].
  \label{eq:nonSM:tensors}
\end{align} 
Generally, $b\to s$ transitions, of which above mentioned process is an example, lead to FCNC which are described in the SM by one-loop diagrams as shown in Fig.\ref{Fig:Feynman B decays}.

After decomposing the hadronic matrix elements for $B_s\to \ell_i^- \ell_j^+$ decay mode into $\langle 0|\bar{b}\gm \gamma_5 s|B_s(p)\rangle =if_{B_s}p_{\mu}$, where $f_{B_s}$ being the $B_s$-meson decay constant, one can obtain the branching fraction containing both vector and scalar kind of operators as \cite{Becirevic:2016zri},
\begin{eqnarray}
\Br[B_s\to \ell_i^- \ell_j^+] &=& \frac{\tau_{B_s}}{64 \pi^3}\frac{\alpha^2 G_F^2}{m_{B_s}^3}  f_{B_s}^2 |V_{tb}V_{ts}^*|^2 \lambda^{1/2}(m_{B_s},m_i,m_j) \nonumber \\
&& 
\times \Bigg{\lbrace}[m_{B_s}^2-(m_i+m_j)^2]\cdot\left|(C_9^{\ell_i\ell_j}-C_9^{\prime\ell_i\ell_j})(m_i-m_j)+(C_S^{\ell_i\ell_j} - C_S^{\prime \ell_i\ell_j})\frac{m_{B_s}^2}{m_b+m_s}\right|^2 \nonumber \\
&&
+[m_{B_s}^2-(m_i-m_j)^2]\cdot\left|(C_{10}^{\ell_i\ell_j}  - C_{10}^{\prime\ell_i\ell_j})(m_i+m_j)+(C_{P}^{\ell_i\ell_j}-C_{P}^{\prime \ell_i\ell_j})\frac{m_{B_s}^2}{m_b+m_s}\right|^2 \Bigg{\rbrace}. \nonumber\\ \label{eq:Bs to l1l2}
\end{eqnarray}
For $B\to K^{(*)}\ell_i^+\ell_j^-$, the branching fraction containing both vector and scalar kind of operators is written as
\begin{eqnarray}
\Br[B\to K^{(*)}\ell_i^+\ell_j^-] &=& 10^{-9} \bigg\{ a_{K^{(*)}\ell_i\ell_j} \left|C_9^{\ell_i\ell_j} + C_9^{\prime\ell_i\ell_j} \right|^2  + b_{K^{(*)}\ell_i\ell_j} \left|C_{10}^{\ell_i\ell_j}  + C_{10}^{\prime\ell_i\ell_j} \right|^2 \nonumber \\ 
&& 
+c_{K^*\ell_i\ell_j}\left|C_9^{\ell_i\ell_j} - C_9^{\prime\ell_i\ell_j} \right|^2 + d_{K^*\ell_i\ell_j}\left|C_{10}^{\ell_i\ell_j}-C_{10}^{\prime\ell_i\ell_j} \right|^2 \nonumber \\
&&
+e_{K^{(\ast)}\ell_i\ell_j} \Big{|}C_S^{\ell_i\ell_j}+C_S^{\prime \ell_i\ell_j}\Big{|}^2 + f_{K^{(\ast)}\ell_i\ell_j} \Big{|}C_{P}^{\ell_i\ell_j}+C_{P}^{\prime \ell_i\ell_j} \Big{|}^2 \nonumber\\
&&
+ g_{K^{(\ast)} \ell_i\ell_j}  \Big{|}C_{S}^{\ell_i\ell_j} -C_{S}^{\prime \ell_i\ell_j}\Big{|}^2+h_{K^{(\ast)} \ell_i\ell_j}  \Big{|}C_{P}^{\ell_i\ell_j}-C_{P}^{\prime \ell_i\ell_j} \Big{|}^2\bigg\}. \label{eq:B to Kstar-l1l2}
\end{eqnarray}

\noindent
In Eq.~\ref{eq:Bs to l1l2}, $\tau_{B_s}$ and $M_{B_s}$  represent the lifetime and mass of $B_s$ particle respectively, $m_{\ell}$s are lepton masses, $\alpha$ is the fine-structure constant and $\lambda (a,b,c)= [a^2 - (b-c)^2][a^2-(b+c)^2] $. In Eq.\ref{eq:B to Kstar-l1l2}, $a_{K^{(*)}\ell_i\ell_j} \cdots h_{K^{(*)}\ell_i\ell_j}$ represent numerical values multiplying the WCs are different for a different choice of flavors \cite{Becirevic:2016zri, Calibbi:2015kma}. The experimental data on such LFV processes in $B$-meson decays are provided by different experimental collaborations as shown in \ref{Tab:LFVlimits}. 
 
                    When the heavy fields are integrated out from the SMEFT, equating the Hamiltonian (Eq.~\ref{eq: Eff Hamiltn}) at $m_W$ scale with the four-fermion currents that are listed in  Table~\ref{tab:BtoS operators1} and keeping the relevant operators only, we get \cite{Alonso:2014csa}
\bea 
C_9^{\ell_i\ell_j} &=&  \dfrac{ (4 \pi)^2 }{e^2 \lambda_{bs}} \frac{v^2}{\Lambda^2}\bigg(\qe^{\ell_i\ell_j} +C_{\ell q}^{(1)\ell_i\ell_j}+C_{\ell q}^{(3)\ell_i\ell_j}\bigg) \label{eq:C9}\,, \\
C_9^{\prime \ell_i\ell_j} &=&  \dfrac{  (4 \pi)^2}{e^2 \lambda_{bs}} \frac{v^2}{\Lambda^2}\bigg (\ed^{\ell_i\ell_j} + \ld^{\ell_i\ell_j} \bigg) \label{eq:C9 prime}\,, \\
C_{10}^{\ell_i\ell_j} &=&  \dfrac{(4 \pi)^2}{e^2 \lambda_{bs}} \frac{v^2}{\Lambda^2}\bigg(\qe^{\ell_i\ell_j} -C_{\ell q}^{(1)\ell_i\ell_j}-C_{\ell q}^{(3)\ell_i\ell_j}\bigg) 
\label{eq:C10}\,, \\
C_{10}^{\prime \ell_i\ell_j} &=&  \dfrac{ (4 \pi)^2 }{e^2 \lambda_{bs}} \frac{v^2}{\Lambda^2}\bigg (\ed^{\ell_i\ell_j} - \ld^{\ell_i\ell_j}  \bigg) \label{eq:C10 prime}
\eea
and similarly for scalar and pseudo-scalar operators
\bea
C_{S}^{\ell_i\ell_j} &=& -C_P^{\ell_i\ell_j} =  \dfrac{ (4 \pi)^2}{e^2 \lambda_{bs}} \frac{v^2}{\Lambda^2} C_{\ell edq}^{\ell_i\ell_j}
\label{eq:CSP}\,,\\
C_{S}^{\prime \ell_i\ell_j} &=& C_P^{\prime \ell_i\ell_j} =  \dfrac{ (4 \pi)^2}{e^2 \lambda_{bs}} \frac{v^2}{\Lambda^2} C_{\ell edq}^{\prime \ell_i\ell_j}
\label{eq:CSPp}\,,
\eea
where the primed operator $C_{\ell edq}^{\prime \ell_i\ell_j}$ represents a different flavor entry of the hermitian conjugate of the corresponding unprimed operator, $\lambda_{bs} = V_{tb}V_{ts}^*$. 

Looking at the structure of the operators $\cO_{9}^{(\prime)}$, $\cO_{10}^{(\prime)}$, $\cO_{S}^{(\prime)}$ and $\cO_{P}^{(\prime)}$ we find that the $2q2\ell$ operators listed in Table~\ref{tab:BtoS operators1} are most relevant for our purpose as we have already mentioned above. The associated WCs of all these operators are to be evaluated at the $b$-quark mass scale $(\mu = m_b)$ as we are interested in LFV processes in $B$-meson decays. Therefore the RGEs of those WCs are needed to run down from some high energy scale and as we have mentioned before, such SMEFT running will induce several other WCs which are not explicitly related to LFVBDs. In the following sections, we shall list all dim-6 SMEFT operators contributing to the processes under our consideration. 


\subsection{Muon to electron conversion in nuclei ($\rm CR (\mu \rightarrow e)$)}
\label{Subsec: CR-mue}

The most general LFV interaction Lagrangian, which
contributes to the $\mu$-$e$ transition in nuclei is given by \cite{Kuno:1999jp}:
\begin{eqnarray}
    {\cal L}_{\rm eff} &=&
    - \frac{4 G_{\rm F}}{\sqrt{2}}
    \left(
    m_\mu A_R \bar{\mu} \sigma^{\mu \nu} P_L e F_{\mu \nu}
    + m_\mu A_L \bar{\mu} \sigma^{\mu \nu} P_R e F_{\mu \nu}
    + {\rm h.c.}
    \right) \nonumber \\
    &&
    - \frac{G_{\rm F}}{\sqrt{2}} \sum_{q = u,d,s}
    \left[ {\rule[-3mm]{0mm}{10mm}\ } \right.
    \left(
    g_{LS}^{(q)} \bar{e} P_R \mu + g_{RS}^{(q)} \bar{e} P_L \mu
    \right) \bar{q} q  \nonumber \\
    &&   \hspace*{2.4cm}
    +
    \left(
    g_{LP}^{(q)} \bar{e} P_R \mu + g_{RP}^{(q)} \bar{e} P_L \mu
    \right) \bar{q} \gamma_5 q \nonumber \\
    &&   \hspace*{2.4cm}
    +
    \left(
    g_{LV}^{(q)} \bar{e} \gamma^{\mu} P_L \mu
    + g_{RV}^{(q)} \bar{e} \gamma^{\mu} P_R \mu
    \right) \bar{q} \gamma_{\mu} q \nonumber \\
    &&   \hspace*{2.4cm}
    +
    \left(
    g_{LA}^{(q)} \bar{e} \gamma^{\mu} P_L \mu
    + g_{RA}^{(q)} \bar{e} \gamma^{\mu} P_R \mu
    \right) \bar{q} \gamma_{\mu} \gamma_5 q \nonumber \\
    &&   \hspace*{2.4cm}
    + \ \frac{1}{2}
    \left(
    g_{LT}^{(q)} \bar{e} \sigma^{\mu \nu} P_R \mu
    + g_{RT}^{(q)} \bar{e} \sigma^{\mu \nu} P_L \mu
    \right) \bar{q} \sigma_{\mu \nu} q
	+ {\rm h.c.}
    \left. {\rule[-3mm]{0mm}{10mm}\ } \right]\ ,
    \label{1}
\end{eqnarray}
where $A_{L,R}$ and $g$'s are all dimensionless coupling constants for the corresponding operators. From this effective Lagrangian the $\mu$-$e$ conversion rate in nuclei can be expressed by the formula ~\cite{Kitano:2002mt},
\be
\Gamma_{\mu \rightarrow e~\rm conv} = \frac{m_\mu^5}{\omega_{\rm capt}\Lambda^4}\bigg\{
\Big| \widetilde C_{DL} D + \widetilde C_{SL}^{(p)} S^{(p)} + \widetilde C_{SL}^{(n)} S^{(n)} + \widetilde C_{VL}^{(p)} V^{(p)} + \widetilde C_{VL}^{(n)} V^{(n)} \Big|^2
+\Big| L\leftrightarrow R \Big|^2 \bigg\}\,,
\label{eq:CRmue BR}
\ee
where $\omega_{\rm capt}$ being the muon capture rate in nuclei $N$ and $D, S^{(p/n)}$, $V^{(p/n)}$ represent dimensionless overlap integrals for dipole, scalar and vector operators respectively. Their numerical values depend on the nuclei and can be found in Ref.~\cite{Kitano:2002mt}.
After tree-level matching~\cite{Jenkins:2017jig}, we obtain that the dipole form factors are given by
\be
 A_L = \frac v{2\sqrt2 m_\mu}\, C_\gamma^{e\mu}\,, \qquad  
 A_R = \frac v{2\sqrt2 m_\mu}\, C_\gamma^{\mu e\, *}\,,
\ee
the vector form factors by
\be
\widetilde C_{VL}^{(p)} = 2  g_{LV,RV}^{(u)} +g_{LV,RV}^{(d)}\,,\qquad
\widetilde C_{VL}^{(n)} =g_{LV,RV}^{(u)} + 2 g_{LV,RV}^{(d)}\,,
\ee
with
\begin{align}
g_{VL}^{(u)} &=  \Big(C_{\ell q}^{(1)} - \lq3 + C_{\ell u}\Big)^{e\mu uu} + (1-\frac83 \sw^2) \Big(C_{\vp l}^{(1)} + C_{\vp l}^{(3)}\Big)^{e\mu}\,,\label{eq:gVLu}\\
g_{VL}^{(d)} &= \Big(C_{\ell q}^{(1)} + \lq3 + \ld\Big)^{e\mu dd} -(1-\frac43 \sw^2) \Big(C_{\vp l}^{(1)} + C_{\vp l}^{(3)}\Big)^{e\mu}\,,\label{eq:gVLd}\\
g_{VR}^{(u)} &= C_{e u}^{e\mu uu} +\qe^{uue\mu} + (1-\frac83 \sw^2) \phie^{e\mu}\,,\label{eq:gVRu}\\
g_{VR}^{(d)} &= C_{e d}^{e\mu dd} +\qe^{dde\mu} - (1-\frac43 \sw^2) \phie^{e\mu}\,,\label{eq:gVRd}
\end{align}
and finally, the scalar form factors by
\begin{align}
\widetilde C_{SL}^{(p/n)} &= - G_S^{(u,p/n)}\, C_{\ell equ}^{(1)\, e\mu uu} + G_S^{(d,p/n)}\, C_{\ell edq}^{e\mu dd} + G_S^{(s,p/n)} C_{\ell edq}^{e\mu ss}\,,\\
\widetilde C_{SR}^{(p/n)} &= - G_S^{(u,p/n)}\, C_{\ell equ}^{(1)\, \mu euu\, *} + G_S^{(d,p/n)}\, C_{\ell edq}^{\mu edd\, *} + G_S^{(s,p/n)} C_{\ell edq}^{\mu ess\, *}\,,\\
\end{align}
with the numerical coefficients~\cite{Kitano:2002mt}
\be
G_S^{(u,p)}=G_S^{(d,n)}=5.1\,,\quad 
G_S^{(u,n)}=G_S^{(d,p)}=4.3\,,\quad
G_S^{(s,p)}=G_S^{(s,n)}=2.5\,.
\ee


\subsection{Lepton flavor violating 3 body leptonic decays  $\boldsymbol{(\ell_i \to \ell_j\ell_k\bar\ell_m})$}
\label{Subsec:3 body decays}

Starting from the LEFT Lagrangian~\cite{Jenkins:2017jig}, the relevant terms for the tree-level 3 body decays are
\begin{align}
\mathcal L_{\rm LEFT} &\supset
C_{ee}^{VLL}\, \big(\bar e_j \gamma^\mu P_L e_i\big) \big(\bar e_k \gamma_\mu P_L e_m\big)
+C_{ee}^{VRR}\, \big(\bar e_j \gamma^\mu P_R e_i\big) \big(\bar e_k \gamma_\mu P_R e_m\big) \nonumber\\
&+ C_{ee}^{VLR}\, \big(\bar e_j \gamma^\mu P_L e_i\big) \big(\bar e_k \gamma_\mu P_R e_m\big)
+ \Big\{ C_{ee}^{SRR} \big(\bar e_j P_R e_i\big)\big(\bar e_k P_R e_m\big) + h.c. \Big\} \nonumber\\
&+ \Big\{ C_\gamma \big(\bar e_j \sigma^{\mu\nu} P_R e_i\big) F_{\mu\nu} + h.c. \Big\}\,.
\end{align} 
The expressions for the decays depend on the flavor combinations of the final leptons, as they could involve new possible contractions and symmetry factors. Therefore, the general expression for the branching ratio of three-body charged lepton decays is given by \cite{Crivellin:2013hpa}
\bea
\mathrm{Br}(\ell_i\to \ell_j \ell_k \bar \ell_l) &=& \frac{N_c
  M^5}{6144\pi^3 \Lambda^4 \Gamma_{\ell_i}} \left(4 \left(|C_{VLL}|^2
+ |C_{VRR}|^2 + |C_{VLR}|^2 + |C_{VRL}|^2\right)\right. 
+|C_{SLL}|^2 \nnu
 &+& |C_{SRR}|^2+  |C_{SLR}|^2+ |C_{SRL}|^2 \left.   48 \left(|C_{TL}|^2 + |C_{TR}|^2\right) + X_\gamma\right)
\label{eq:br4l}
\eea
where $N_c=1/2$ if two of the final state leptons are identical, $N_c=1$ in all other cases and $\Gamma_{\ell_i}$ is the total decay width of the initial lepton. Due to the hierarchy of the charged lepton masses, it is assumed that $m_i\equiv M\gg m_j,m_k,m_l$ and the lighter lepton masses are neglected. $C_X$ are different for different processes.
For decay of the type $\ell_i\to \ell_j\ell_j\bar \ell_j$
\begin{align}
C_{VLL} &= 2\left( (2s_W^2-1) \left( C_{\vp \ell}^{(1)ji} + C_{\vp
  \ell}^{(3)ji} \right) + C_{\ell\ell}^{jijj} \right)\,, \label{eq:leptonic-3-body-decays-1}\\
C_{VRR} &= 2 \left( 2 s_W^2 \phie^{ji} +
C_{ee}^{jijj} \right)\,,\label{eq:leptonic-3-body-decays-2}\\
C_{VLR} &= -\frac{1}{2} C_{SRL} =  2s_W^2 \left( C_{\vp
  \ell}^{(1)ji} + C_{\vp \ell}^{(3)ji} \right) + C_{\ell e}^{jijj}\,,\label{eq:leptonic-3-body-decays-3}\\
C_{VRL} &= -\frac{1}{2} C_{SLR} =  (2s_W^2-1) \phie^{ji} +
C_{\ell e}^{jjji}\,,\label{eq:leptonic-3-body-decays-4} \\
C_{SLL} &= C_{SRR} =C_{TL} =C_{TR} = 0\,, \\
C_{\gamma L}&= \sqrt{2}C_\gamma^{ij\star}\,, \\
C_{\gamma R}&=  \sqrt{2}C_\gamma^{ji}\,,
\end{align}
where $C_\gamma^{ji}$, known as the photon dipole operator, is defined as 
\be
 C_\gamma^{fi} \equiv \left(c_W C_{eB}^{fi} - s_W
C_{eW}^{fi} \right) \,
\label{eq:Cgamma}
\ee
and
\bea
X_\gamma & =& -\frac{16ev}{M}\mathrm{Re}\left[\left(2 C_{VLL} +
  C_{VLR} - \frac{1}{2} C_{SLR} \right) C_{\gamma R}^\star + \left( 2
  C_{VRR} + C_{VRL} - \frac{1}{2} C_{SRL} \right) C_{\gamma L}^\star
  \right] \nnu
&+& \frac{64 e^2v^2}{M^2} \left(\log\frac{M^2}{m^2} - \frac{11}{4}
\right)(|C_{\gamma L}|^2 + |C_{\gamma R}|^2) \,.
\eea


\subsection{Operators relevant for different LFV  processes}

\begin{table}[hbt!] 
\centering
\renewcommand{\arraystretch}{1.5}
\begin{tabular}{|c|c|} 
\hline 
Processes & Most relevant operators  \\
\hline \hline
$B\rightarrow K\ell_i\ell_j$ & $\cO_{\ell q}^{(1)}, \cO_{\ell q}^{(3)}, \cO_{qe}, \cO_{\ell d}, \cO_{ed}$, $\cO_{\ell edq}$   \\
\hline
$B\rightarrow K^*\ell_i\ell_j$ & $\cO_{\ell q}^{(1)}, \cO_{\ell q}^{(3)}, \cO_{qe}, \cO_{\ell d}, \cO_{ed}$, $\cO_{\ell edq}$  \\
\hline
$B_s\rightarrow \mu e$ & $\cO_{\ell q}^{(1)}, \cO_{\ell q}^{(3)}, \cO_{qe}, \cO_{\ell d}, \cO_{ed}$, $\cO_{\ell edq}$   \\
\hline \hline
$\ell_i\rightarrow \ell_j\gamma$  & $\cO_{eB}, \cO_{eW}$  \\
\hline\hline
$\ell_i\rightarrow \ell_j\ell_j\bar{\ell}_j$  & $\cO_{\vp \ell }^{(1)},  \cO_{\vp \ell }^{(3)}, \cO_{\vp \e }, \cO_{\ell\ell }, \cO_{\ell e}, \cO_{e e}$   \\
\hline 
$\ell_i\rightarrow \ell_j\ell_k\bar{\ell}_k$  & $\cO_{\vp \ell }^{(1)},  \cO_{\vp \ell }^{(3)}, \cO_{\vp \e }, \cO_{\ell\ell }, \cO_{\ell e}, \cO_{e e}$  \\
\hline \hline
$\rm CR(\mu\rightarrow e)$ & $ \cO_{\vp \ell }^{(1)}, \cO_{\vp \ell }^{(3)}, \cO_{\vp e }, \cO_{e u}, \cO_{\ell u}, \cO_{\ell q}^{(1)}, \cO_{\ell q}^{(3)}, \cO_{q e}, \cO_{\ell d}, \cO_{e d}, \cO_{\ell edq}, \cO_{\ell equ}$  \\
\hline \hline
$Z\rightarrow \ell_i\ell_j$ & $\cO_{\vp \ell }^{(1)},  \cO_{\vp \ell }^{(3)}, \cO_{\vp \e }, \cO_{eB}, \cO_{eW}$ \\
\hline\hline
$\tau \to \mesonV \ell$ ($\mesonV=\rho, \phi$) & $\cO_{\vp \ell }^{(1)},  \cO_{\vp \ell }^{(3)}, \cO_{\vp \e },  \cO_{\ell u}, \cO_{eu}, \cO_{\ell equ}, \cO_{eB}, \cO_{eW}, \cO_{\ell q}^{(1)}, \cO_{\ell q}^{(3)}, \cO_{qe}, \cO_{\ell d},  \cO_{ed}$ \\
\hline
$\tau \to \mesonP \ell$ ($\mesonP=\pi^0, K^0$) & $\cO_{\vp \ell }^{(1)},  \cO_{\vp \ell }^{(3)}, \cO_{\vp \e },  \cO_{\ell u}, \cO_{eu}, \cO_{\ell equ}, \cO_{eB}, \cO_{eW}, \cO_{\ell q}^{(1)}, \cO_{\ell q}^{(3)}, \cO_{qe}, \cO_{\ell d}, \cO_{ed}, \cO_{\ell edq}$ \\
\hline
\end{tabular}
\caption{List of the dimension-$6$ operators (invariant under the SM gauge group) which contribute to different LFV processes under consideration at the tree or at the one-loop level.
\label{tab:LFV Oprtr}}
\end{table}

In the model-independent approach of EFT, we either obtain the experimental constraints on the WCs or determine the non-zero value of those coefficients if there is an SM value-deviating signal. In order to do so, the renormalization group equations of the WCs of these relevant operators, represented as the combinations of anomalous-dimension $(\gamma_{ij})$ matrix of these dimension-six operators, need to be solved. These anomalous dimensions are defined as 
\be \label{eq:anom dimsn}
\dot{C_i}\equiv 16\pi^2\mu \frac{d C_i}{d\mu}=\gamma_{ij}C_j.
\ee
Since flavor-changing effects are propagated through the Yukawa RGEs  \cite{Jenkins:2013wua}, we shall consider them in detail. There will be an additional 30 SMEFT operators, besides 6 primary operators, that contribute to LFVBD processes in different strengths. Therefore it is useful to represent their contributions in a schematic form as 
\be
\left(\begin{array}{c} \dot \Cone \\ \dot \Ctwo \\ \dot \Cthree \\ \dot\Cfour 
\end{array}\right)_Y \equiv
16\pi^2 \mu \frac d{d\mu} \left(\begin{array}{c}  \Cone\\  \Ctwo \\  \Cthree \\  \Cfour 
\end{array}\right)_Y =
\left(\begin{array}{cccc}
\gamma_{11} & \gamma_{12} & 0 & 0  \\
\gamma_{21} & \gamma_{22} & \gamma_{23} & 0  \\
0 & \gamma_{32} & \gamma_{33} & \gamma_{34}  \\ 
0 & 0 & \gamma_{43} & \gamma_{44}  \\
\end{array}\right)_Y\, 
\left(\begin{array}{c}  \Cone\\  \Ctwo \\  \Cthree \\  \Cfour 
\end{array}\right)_Y \,,
\label{eq:RGEmat}
\ee
with $Y$ denoting the Yukawa RGEs and
\begin{align} 
(\Cone)_Y \equiv& \Big( \lQ1, \lq3, \qe, \ld, \ed, C_{\ell edq} \Big)^T\,,\label{eq:C1}\\
(\Ctwo)_Y \equiv& \Big( C_{\vp q}^{(1)}, C_{\vp q}^{(3)}, \Phil1, \phil3, C_{\vp d}, \phie, C_{\ell u}, C_{\ell e}, C_{eu},  C_{qq}^{(1)}, C_{qq}^{(3)}, C_{qd}^{(1)}, C_{qd}^{(8)}, \nnu 
& C_{\ell equ}^{(1)}, C_{\ell equ}^{(3)}, C_{quqd}^{(1)},  C_{quqd}^{(8)}\Big)^T\,, \label{eq:C2} \\ 
(\Cthree)_Y \equiv & \Big(C_{\vp\framebox(4,4){}}, C_{\vp D}, C_{\vp u}, C_{\vp d}, C_{\vp e}, C_{\vp ud},  C_{ud}^{(1)}, C_{qu}^{(1)}, C_{qu}^{(8)}, C_{dd}, C_{\ell \ell}, C_{ee}\Big)^T\,, \label{eq:C3} \\
(\Cfour)_Y \equiv & \Big( C_{uu} \Big)^T\,, \label{eq:C4}
\end{align}
and $(\gamma_{ij})_Y$ represent the anomalous dimensions whose explicit forms, in terms of other WCs, are given in Ref.~\cite{Alonso:2013hga}. The form of $\gamma$ matrix, in Eq.~\ref{eq:RGEmat} tells us that RGEs of WCs of primary operators, mainly responsible for LFVBD and given by Eq.~\ref{eq:C1}, depend on themselves and 17 other WCs listed in  $\Ctwo$ (Eq.~\ref{eq:C2}). These in turn depend on 12 other WCs besides themselves and similarly the single WC, $C_{uu}$ in $\Cfour$ completes the list of all WCs involved in LFVBD processes. These dependencies of WCs or operators are listed in $\Ctwo$, $\Cthree$ and $\Cfour$ respectively and thus $\gamma$ matrix elements bear non-zero contributions for all elements except  $\gamma_{13}$ and $\gamma_{14}$ as none of the operators listed in Eq.~\ref{eq:C1} depend on operators in Eq.~\ref{eq:C4}. Table~\ref{tab:LFV Oprtr} represents the list of most relevant operators, or `Primary Operators', responsible for LFV processes under consideration in this analysis. In a similar fashion as described above, one can arrange and formulate the gamma functions for all direct and induced operators for a particular LFV process.


\section{Results and Discussion}
\label{Sec:Results}

  We focus on probing new physics scenarios in a model-independent framework like SMEFT while primarily considering cLFV processes in B decays. Apart from the operators that directly affect the LFVBDs, we also consider a few relevant WCs associated with other LFV processes. We enumerate the operators associated with all the LFV processes considered in this analysis in
  Table \ref{tab:LFV Oprtr}. 
As mentioned earlier we compute the BRs of charged lepton flavor violating observables using the package {\it flavio}~\cite{Straub:2018kue}. The package 
{\it wilson}~\cite{Aebischer:2018bkb} is used for the RGE running of the WCs. We remind that all the constraints used in this analysis (Table~\ref{Tab:LFVlimits}) come from the upper limits of experimental data of the associated LFV processes rather than any explicit measurements having non-vanishing central values along with error estimates.

  In the first part, we select the SMEFT WCs that are important for LFVBDs and probe the energy scale $\Lambda$ from the experimental bounds of all the LFV processes. The associated WCs considered one at a time at the scale $\Lambda$,
are set to unity, whereas all the other WCs are set to zero at the same scale. Certainly, this involves i) RGE running and the match and run procedure involving the SMEFT and the LEFT operators, as described earlier, and ii) finding the low energy observables for LFVBDs at the scale of $m_b$. The other LFV processes involve further smaller scales like $m_\tau$ or $m_\mu$. We also remind ourselves that various WCs vanishing at the scale 
$\Lambda$ may receive finite contributions at $m_W$ due to RGE mixing. In this context, we will estimate the relative strengths of WCs in Table~\ref{Tab:correlation}.

  Going a step further, considering a fixed
  value of $\Lambda$, we will also consider several cases of two 
  non-vanishing WCs and study the effects of the combined LFV constraints.  
  We note that while considering a given value of
  BR of a process, there may be three broad situations: 
  i) contributions from one of the WCs may be negligible; ii) the same 
  may be comparable to each other, and they add up constructively, and iii)
  the same may interfere destructively.
  Plotted on a logarithmic scale, the case (i) would result in contours
  almost parallel to one of the WC axes; case (ii) would produce round curves,
  whereas case (iii) would show up as cuspy regions.  
  Clearly, in the last case, one requires larger absolute values of both the
  WCs and the associated regions correspond to the so-called
  {\it flat-directions}.


\subsection{Single Operator Analysis}

As mentioned earlier, in this part we choose a single WC initialized to unity at the SMEFT scale of $\mu=\Lambda$ and evolve it toward a low-scale $m_W$ via appropriate RGEs. We will set all other WCs to zero at $\Lambda$. Based on our earlier discussion on LFVBDs, we have the following single operator scenarios where WCs of primary importance are: $[C^{(1,3)}_{\ell q}]_{\ell\ell'23}$, $[\qe]_{\ell\ell'23}$,
$[\ed]_{\ell\ell'23}$, $[\ld]_{\ell\ell'23}$ and $[\ledq]_{\ell\ell'23}$. Here, $\ell\ell' \in (12,13,23)$ represents the combinations $e-\mu$, $e-\tau$ and $\mu-\tau$ respectively. At the low energy scale, the first three WCs namely, $[C^{(1)}_{\ell q}]_{\ell\ell'23}, [C^{(3)}_{\ell q}]_{\ell\ell'23}$ and $[\qe]_{\ell\ell'23}$, contribute to left-handed $C_{9,10}$, the next two contribute to right-handed $C'_{9,10}$, and the last one and its chiral 
counterpart  relate to $C_{S,P}$ and $C^{\prime}_{S,P}$ of LFVBDs respectively (Eqs.~\ref{eq:C9} to \ref{eq:CSPp}).
 
 As indicated by the RGEs (Eqs. \ref{eq:anom dimsn}-\ref{eq:C4})
  WCs of different operators mix among themselves, and a given vanishing
  WC at high-energy scale may accumulate a non-zero value at a low-energy scale like $m_W$. Furthermore, it is necessary to explore the mutual dependence of the 
  primarily important WCs associated with the LFVBDs, and a few of the WCs important for the other LFV
  processes due to RGE effects. Thus, we choose to study the 
  above effects of all the WCs grouped under ${\Cone}_Y$ (Eq.~\ref{eq:C1}) 
  and $\Phil1$, $\phil3$ and $\phie$ of ${\Ctwo}_Y$ (Eq. \ref{eq:C2})
  that are relevant from Table~\ref{tab:LFV Oprtr}. This leads to   
  Table~\ref{Tab:correlation} where we show the depiction of order of magnitudes of the said WCs due to RGE effects, while ignoring signs. 
  We use $\Lambda=1$~TeV and set the desired non-vanishing
  WC to unity at the same scale, indicating a coupling $(C/\Lambda^2)$ value of  $10^{-6} \GeV^{-2}$, here $C$ refers to a given WC.
  The first 9 rows and 9 columns of Table~\ref{Tab:correlation} show the
  interdependence of the said WCs.
  Additionally, from the 10th row onward, we explore 
  the effects on a few more WCs that are relevant in the context of the other LFV processes (Table~\ref{tab:LFV Oprtr}). All coupling entries in the table are obtained at the energy scale 
  $\mu = m_W$. The dark grey boxes with identical row and column indices refer to the WCs that are non-vanishing at the higher scale and prominent at $m_W$.
  Tiny coupling values below or equal to $10^{-14}$ appear in white boxes, whereas the values that are relatively prominent and above this limit are shown in light grey boxes.
For example, focusing on the $\lQ1$ column,
the RGE effects have significant 
impacts on $\lq3$, $\Phil1$ and $\phil3$. However, the same on the rest of the WCs 
  is hardly significant. 
 A similar behavior 
  holds true for $\lq3$ as well. Moreover, both $\lQ1$ and $\lq3$ can also affect the dipole operators $C_{eW}$ and $C_{eB}$ which are quite important in the context of the stringent limit from $\Mueg$. Considering $\qe$, we find that it has a relatively prominent RGE impact on $\phie$, 
   whereas for $C_{ledq}$ the scenario is little different. In this case, the relevant WC (with specific quark indices) is $[C_{ledq}]_{1223}$ which has significant impact only on $[C_{ledq}]_{1222}$ (i.e. with different quark indices).

\begin{table}[htbp]
\begin{center}
\resizebox{\textwidth}{!}{
\renewcommand{\arraystretch}{1.3}
\begin{tabular}{|c|c|c|c|c|c|c|c|c|c|c|}
\hline
WCs & $[\lQ1]_{1223}$ & $[\lq3]_{1223}$ & $[\qe]_{2312}$ & $[\ed]_{1223}$ & $[\ld]_{1223}$ & $[\ledq]_{1223}$ & $[\Phil1]_{12}$ & $[\phil3]_{12}$ & $[\phie]_{12}$  \\
\hline 
\hline
$[\lQ1]_{1223}$  & \cellcolor{gray!40}$10^{-6}$ & \cellcolor{gray!20}$10^{-8}$ & $10^{-17}$ & $10^{-24}$ & $10^{-14}$ & $10^{-16}$ & \cellcolor{gray!20}$10^{-10}$ & \cellcolor{gray!20}$10^{-11}$ & $10^{-20}$ \\
\hline
$[\lq3]_{1223}$ & \cellcolor{gray!20}$10^{-8}$ & \cellcolor{gray!40}$10^{-6}$ & $10^{-19}$ & $10^{-25}$ & $10^{-16}$ & $10^{-16}$ & \cellcolor{gray!20}$10^{-12}$ & \cellcolor{gray!20}$10^{-9}$ & $10^{-23}$ \\
\hline
$[\qe]_{2312}$ & $10^{-17}$ & $10^{-18}$ & \cellcolor{gray!40}$10^{-6}$ & $10^{-14}$ & $10^{-24}$ & $10^{-18}$ & $10^{-20}$ & $10^{-22}$ & \cellcolor{gray!20}$10^{-10}$ \\
\hline
$[\ed]_{1223}$ & $10^{-23}$ & $10^{-23}$ & $10^{-13}$ & \cellcolor{gray!40}$10^{-6}$ & $10^{-17}$ & $10^{-16}$ & $10^{-27}$ & $10^{-27}$ & $10^{-17}$ \\
\hline
$[\ld]_{1223}$ & $10^{-13}$ & $10^{-15}$ & $10^{-24}$ & $10^{-17}$ & \cellcolor{gray!40}$10^{-6}$ & $10^{-14}$ & $10^{-17}$ & $10^{-19}$ & $10^{-27}$ \\
\hline
$[\ledq]_{1223}$ & $10^{-15}$ & $10^{-14}$ & $10^{-17}$ & $10^{-15}$ & $10^{-13}$ & \cellcolor{gray!40}$10^{-6}$ & $10^{-18}$ & $10^{-18}$ & $10^{-21}$ \\
\hline
$[\Phil1]_{12}$ & \cellcolor{gray!20}$10^{-9}$ & \cellcolor{gray!20}$10^{-11}$ & $10^{-20}$ & $10^{-26}$ & $10^{-16}$ & $10^{-18}$ & \cellcolor{gray!40}$10^{-6}$ & \cellcolor{gray!20}$10^{-12}$ & $10^{-17}$\\
\hline
$[\phil3]_{12}$ & \cellcolor{gray!20}$10^{-11}$ & \cellcolor{gray!20}$10^{-9}$ & $10^{-22}$ & $10^{-27}$ & $10^{-19}$ & $10^{-18}$ & \cellcolor{gray!20}$10^{-12}$ & \cellcolor{gray!40}$10^{-6}$ & $10^{-23}$ \\
\hline
$[\phie]_{12}$ & $10^{-19}$ & $10^{-21}$ & \cellcolor{gray!20}$10^{-9}$ & $10^{-16}$ & $10^{-26}$ & $10^{-20}$ & $10^{-17}$ & $10^{-22}$ & \cellcolor{gray!40}$10^{-6}$ \\
\hline
\hline
$[C_{\ell \ell}]_{1112}$ & $10^{-14}$ & $10^{-15}$ & $10^{-22}$ & $10^{-30}$ & $10^{-22}$ & $10^{-21}$ & \cellcolor{gray!20}$10^{-10}$ & \cellcolor{gray!20}$10^{-9}$ & $10^{-20}$ \\
\hline
$[C_{\ell e}]_{1112}$ & $10^{-20}$ & $10^{-19}$ & $10^{-15}$ & $10^{-24}$ & $10^{-27}$ & $10^{-20}$ & $10^{-20}$ & $10^{-20}$ & \cellcolor{gray!20}$10^{-10}$ \\
\hline
$[C_{\ell e}]_{1211}$ & $10^{-15}$ & $10^{-14}$ & $10^{-20}$ & $10^{-27}$ & $10^{-24}$ & $10^{-21}$ & \cellcolor{gray!20}$10^{-9}$ & \cellcolor{gray!20}$10^{-12}$ & $10^{-20}$ \\
\hline
$[C_{ee}]_{1112}$ & $10^{-22}$ & $10^{-22}$ & $10^{-14}$ & $10^{-23}$ & $10^{-30}$ & $10^{-23}$ & $10^{-20}$ & $10^{-22}$ & \cellcolor{gray!20}$10^{-9}$ \\
\hline
$[C_{eu}]_{1211}$ & $10^{-22}$ & $10^{-22}$ & $10^{-15}$ & $10^{-24}$ & $10^{-31}$ & $10^{-24}$ & $10^{-20}$ & $10^{-22}$ & \cellcolor{gray!20}$10^{-10}$ \\
\hline
$[C_{\ell u}]_{1211}$ & $10^{-15}$ & $10^{-15}$ & $10^{-22}$ & $10^{-30}$ & $10^{-24}$ & $10^{-21}$ & \cellcolor{gray!20}$10^{-10}$ & \cellcolor{gray!20}$10^{-12}$ & $10^{-20}$ \\
\hline
$[\lQ1]_{1211}$ & $10^{-15}$ & $10^{-15}$ & $10^{-23}$ & $10^{-30}$ & $10^{-22}$ & $10^{-21}$ & \cellcolor{gray!20}$10^{-10}$ & \cellcolor{gray!20}$10^{-11}$ & $10^{-21}$ \\
\hline
$[\lq3]_{1211}$ & $10^{-14}$ & $10^{-15}$ & $10^{-23}$ & $10^{-30}$ & $10^{-22}$ & $10^{-21}$ & \cellcolor{gray!20}$10^{-13}$ & \cellcolor{gray!20}$10^{-9}$ & $10^{-23}$ \\
\hline
$[\qe]_{1112}$ & $10^{-23}$ & $10^{-23}$ & $10^{-15}$ & $10^{-22}$ & $10^{-31}$ & $10^{-24}$ & $10^{-20}$ & $10^{-21}$ & \cellcolor{gray!20}$10^{-10}$ \\
\hline
$[\ed]_{1211}$ & $10^{-23}$ & $10^{-22}$ & $10^{-15}$ & $10^{-19}$ & $10^{-29}$ & $10^{-24}$ & $10^{-20}$ & $10^{-23}$ & \cellcolor{gray!20}$10^{-10}$ \\
\hline
$[\ld]_{1211}$ & $10^{-15}$ & $10^{-15}$ & $10^{-23}$ & $10^{-29}$ & $10^{-19}$ & $10^{-21}$ & \cellcolor{gray!20}$10^{-10}$ & \cellcolor{gray!20}$10^{-12}$ & $10^{-20}$ \\
\hline
$[\ledq]_{1211}$ & $10^{-19}$ & $10^{-18}$ & $10^{-21}$ & $10^{-25}$ & $10^{-23}$ & $10^{-16}$ & $10^{-19}$ & $10^{-18}$ & $10^{-22}$ \\
\hline
$[\ledq]_{1222}$ & $10^{-19}$ & $10^{-17}$ & $10^{-23}$ & $10^{-19}$ & $10^{-16}$ & \cellcolor{gray!20}$10^{-9}$ & $10^{-17}$ & $10^{-17}$ & $10^{-21}$ \\
\hline
$[C_{\ell equ}^{(1)}]_{1211}$ & $10^{-19}$ & $10^{-19}$ & $10^{-21}$ & $10^{-28}$ & $10^{-26}$ & $10^{-19}$ & $10^{-19}$ & $10^{-19}$ & $10^{-22}$ \\
\hline
$[C_{eB}]_{12}$ & $10^{-15}$ & $10^{-14}$ & $10^{-17}$ & $10^{-25}$ & $10^{-23}$ & $10^{-17}$ & $10^{-16}$ & $10^{-15}$ & $10^{-18}$ \\
\hline
$[C_{eB}]_{21}$ & $10^{-17}$ & $10^{-17}$ & $10^{-15}$ & $10^{-23}$ & $10^{-25}$ & $10^{-26}$ & $10^{-18}$ & $10^{-18}$ & $10^{-16}$ \\
\hline
$[C_{eW}]_{12}$ & $10^{-15}$ & $10^{-14}$ & $10^{-17}$ & $10^{-25}$ & $10^{-22}$ & $10^{-17}$ & $10^{-16}$ & $10^{-15}$ & $10^{-18}$ \\
\hline
$[C_{eW}]_{21}$ & $10^{-17}$ & $10^{-17}$ & $10^{-15}$ & $10^{-22}$ & $10^{-25}$ & $10^{-26}$ & $10^{-18}$ & $10^{-18}$ & $10^{-16}$ \\
\hline
\end{tabular}}
\caption{
Order of magnitude display of RG evolved coupling strengths   
    obtained at the energy scale $\mu=m_W$ for $\Lambda=1$~TeV. 
    The dark grey boxes with identical row and column indices refer to the WCs that are non-vanishing at the higher scale and prominent at $m_W$.
  Tiny coupling values below or equal to $10^{-14}$ appear in white boxes, whereas the values that are relatively prominent and above this limit are shown in light grey boxes.}    
\label{Tab:correlation}
\end{center}
\end{table}

\begin{figure}[htbp]
 \centering
 \begin{subfigure}[b]{0.3\textwidth}
         \centering
         \includegraphics[width=6cm,height=5.0cm]{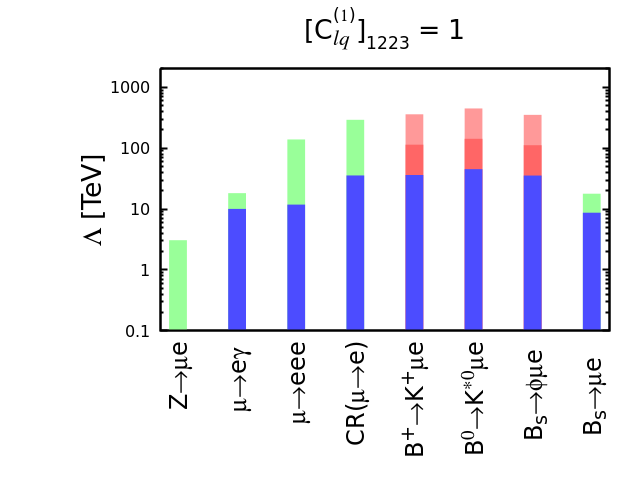}
         \label{Fig:Clq1-1D}
         \caption{}
 \end{subfigure}
 \hfill
 \begin{subfigure}[b]{0.3\textwidth}
         \centering
         \includegraphics[width=6cm,height=5.0cm]{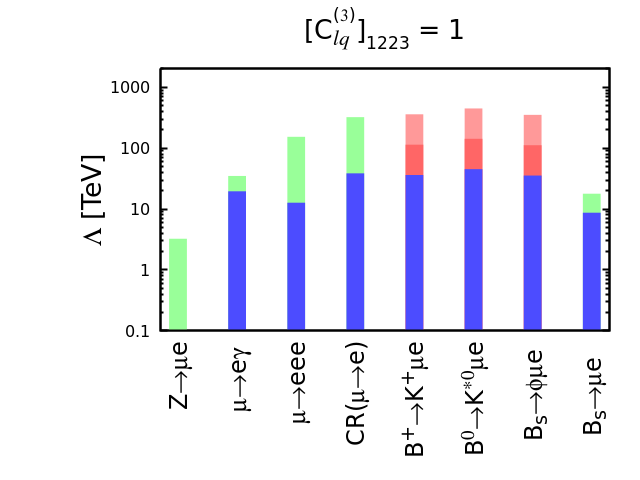}
         \label{Fig:Clq3-1D}
         \caption{}
 \end{subfigure}
 \hfill
 \begin{subfigure}[b]{0.3\textwidth}
         \centering
         \includegraphics[width=6cm,height=5.0cm]{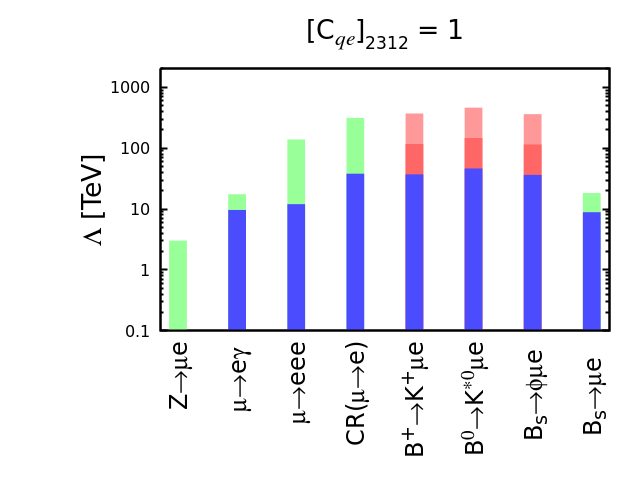}
         \label{Fig:Cqe-1D}
         \caption{}
 \end{subfigure}
 \hfill
 \begin{subfigure}[b]{0.4\textwidth}
         \centering
         \includegraphics[width=6cm,height=5.0cm]{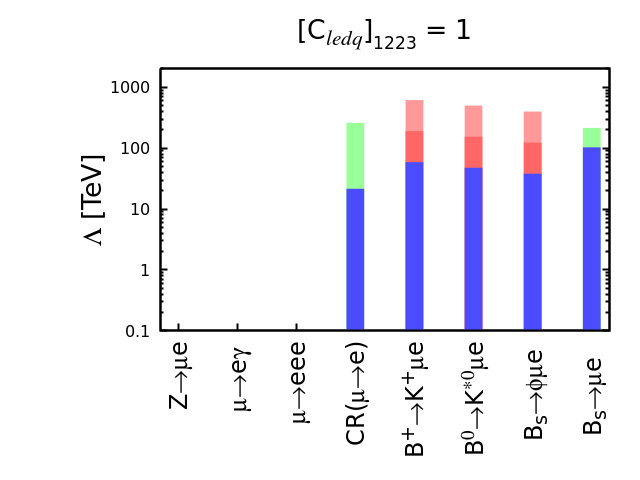}
         \label{Fig:Cledq1-1D}
         \caption{}
 \end{subfigure}
 \begin{subfigure}[b]{0.4\textwidth}
         \centering
         \includegraphics[width=6cm,height=5.0cm]{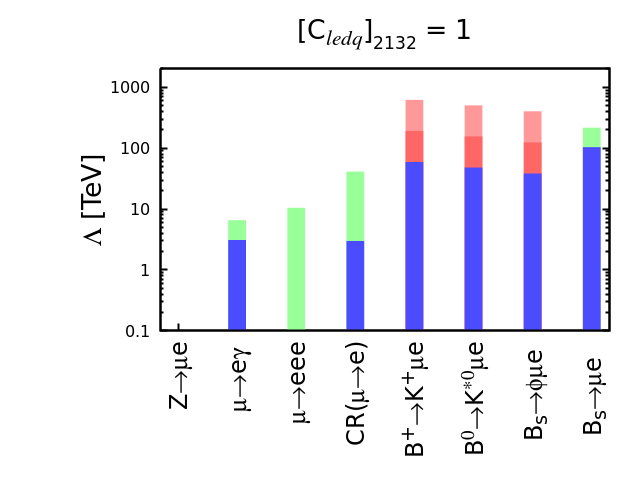}
         \label{Fig:Cledq2-1D}
         \caption{}
 \end{subfigure}
\caption{Display of values of $\Lambda$ for (a) $\lQ1$, (b) $\lq3$, (c) $\qe$ (in the top panel) and (d) \& (e) $\ledq$ (in the bottom panel), for $\mu-e$, that are consistent with the present and future experimental bounds of various LFV decays when only one single (perturbative) WC is fixed at unity. Blue and green bars refer to current and possible future bounds of LFV processes respectively as described in Table~\ref{Tab:LFVlimits}. Darker and lighter red shades, starting above the blue shades, represent the assumed enhancement of the BR sensitivity by 2 and 4 orders of magnitude respectively.}
 \label{Fig:Single-Opr-Dom-1}
 \end{figure}

\begin{figure}[htbp]
 \centering
  \begin{subfigure}[b]{0.3\textwidth}
         \centering
         \includegraphics[width=6cm,height=5.0cm]{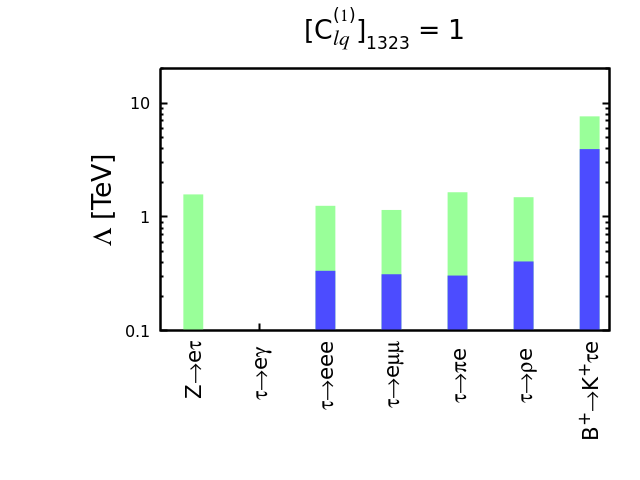}
          \caption{}
         \label{}
 \end{subfigure}
 \hfill
 \begin{subfigure}[b]{0.3\textwidth}
         \centering
         \includegraphics[width=6cm,height=5.0cm]{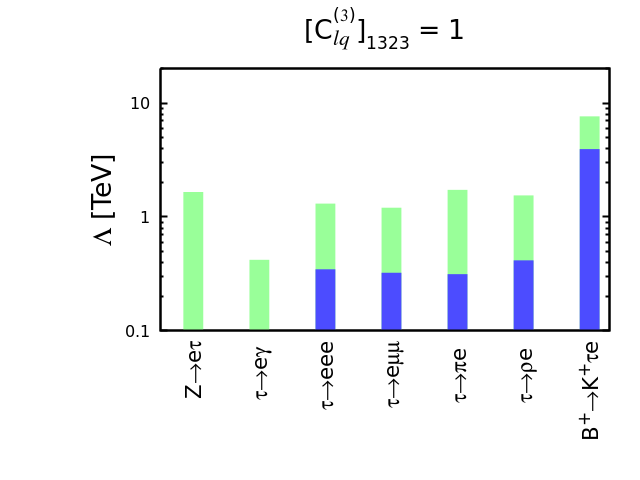}
          \caption{}
         \label{}
 \end{subfigure}
 \hfill
 \begin{subfigure}[b]{0.3\textwidth}
         \centering
         \includegraphics[width=6cm,height=5.0cm]{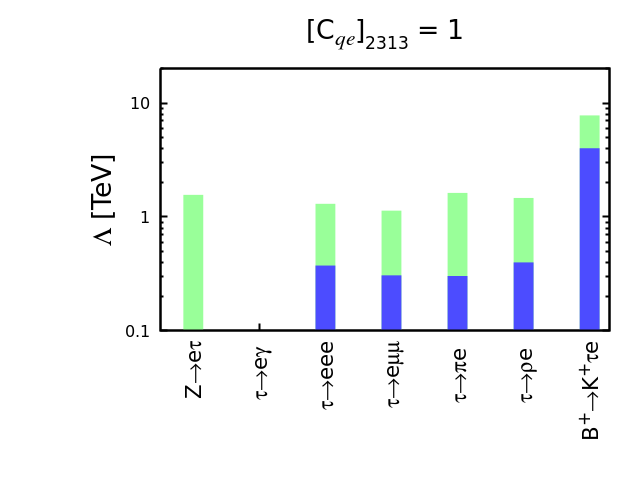}
          \caption{}
         \label{}
 \end{subfigure}
 \hfill
 \begin{subfigure}[b]{0.3\textwidth}
         \centering
          \includegraphics[width=6cm,height=5.0cm]{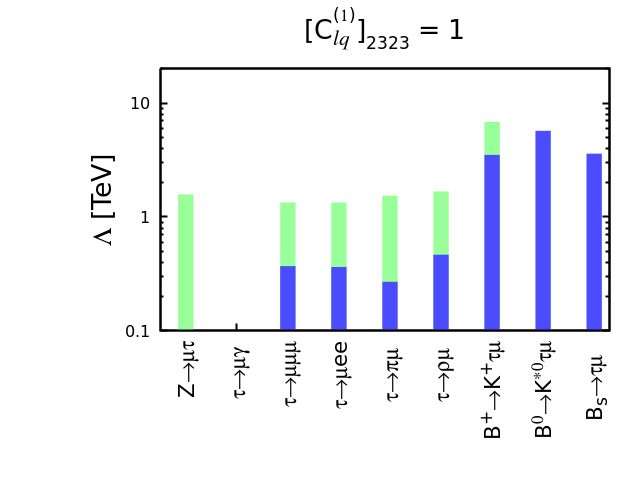}
           \caption{}
         \label{}
 \end{subfigure}
 \hfill
 \begin{subfigure}[b]{0.3\textwidth}
         \centering
         \includegraphics[width=6cm,height=5.0cm]{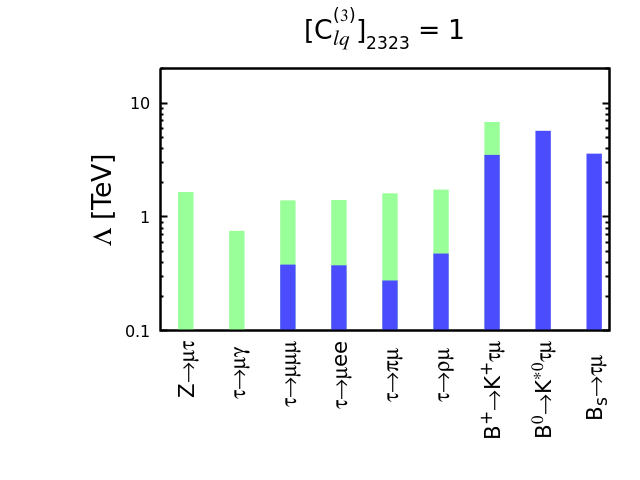}
          \caption{}
         \label{}
 \end{subfigure}
 \hfill
 \begin{subfigure}[b]{0.3\textwidth}
         \centering
         \includegraphics[width=6cm,height=5.0cm]{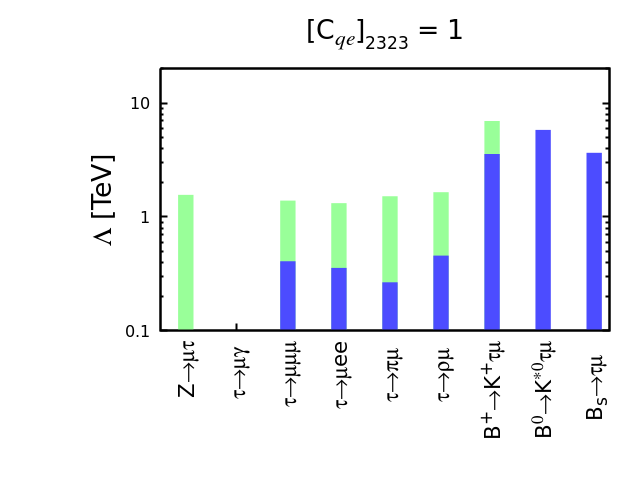}
         \caption{}
         \label{}
 \end{subfigure}
  \hfill
 
 \caption{Similar as Fig.~\ref{Fig:Single-Opr-Dom-1} for $e-\tau$ (in the top panel) and $\mu-\tau$ (in the bottom panel)}
 \label{Fig2:Single-Opr-Dom-1}
 \end{figure}

In Figures \ref{Fig:Single-Opr-Dom-1} and \ref{Fig2:Single-Opr-Dom-1} , we probe the maximum attainable energy
scales denoted as $\Lambda$ by considering the sensitivities of operators relevant to LFV processes (Table~\ref{Tab:LFVlimits}).
We consider the current experimental limits to probe $\Lambda$ and see any prospective change in conclusion by
using possible future bounds of the same processes. We choose the operators that directly affect the LFVBD processes
and constrain $\Lambda$ by those, as well as via other LFV limits. We vary a coupling by changing the scale
$\Lambda$ while fixing the associated WC to unity at that scale. The x-axes in these bar-diagrams display the BRs associated
with different LFV processes, while the y-axes represent the requisite energy scale $\Lambda$ maximally consistent
with the BRs from the table. For a given operator, each LFV process may at most be shown via a bar with two distinct colors: blue color within the bar indicating the current experimental bound on the BRs, and the green color representing the anticipated future constraint on the same as mentioned in the table. 
Among the LFVBDs, the processes which have definite
future sensitivity predictions for BRs like
$\Bsmue$, the color codes continue to remain the same. 
Additionally, we extend our studies to probe how robust would be our 
conclusions in presence of higher degree of sensitivities for some of the
LFVBDs like $\BK$ or $\BksZ$. Accordingly, we use
darker and lighter shadings of red to correspond to sensitivity
enhancement by two and four orders of magnitude respectively.
This extended analysis is indeed performed in the 
same spirit of the recent work of Ref. \cite{Calibbi:2022ddo}. The above 
levels of sensitivity of the mentioned LFVBDs place the latter 
in the same footing as with the predicted improvements of BRs of the 
other LFV processes. 

 In the top panel of Figure \ref{Fig:Single-Opr-Dom-1} we choose three WCs namely, $\lQ1$, $\lq3$, and $\qe$ that are most relevant for LFVBDs, as well as that can influence other LFV processes significantly, although indirectly.
On the other hand, $\ed$ and $\ld$ do not have enough RGE running effects to influence on the other LFV processes, hence these are omitted in this study.
As shown in Figure \ref{Fig:Single-Opr-Dom-1} the LFVBDs like $\cB(B^+\to K^+\mu e)$, followed by 
$\cB(B^0 \to K^{*0} \mu e)$ and $\cB(B_s\to\phi\mu e)$ 
have similar values of $\Lambda$, thus they are quite equally constraining. On the other hand, the effect from 
$\mathcal{B}(B_s \to \mu e)$ is quite less significant. Among other LFV decays, $\CRmue$ and followed by $\Mueee$ provide the most significant constraints to these WCs.
Additionally, the ability of $Z \to \mu e$ to constrain these sets of
WCs is hardly of any significance. Similarly in the bottom panel of Figure \ref{Fig:Single-Opr-Dom-1}, for $\ledq$ (considering both flavor indices 1223 and 2132), we observe that the current constraints predominantly originate from LFVBDs, particularly from processes such as $\mathcal{B}(B_s \to \mu e)$, rather than from $\mathcal{B}(B\to K^{(*)}\mu e)$. This distinction arises due to the varying combinations of pre-factors associated with $\ledq$, a fact that can be verified by referring to Eqs.~\ref{eq:Bstomue1} and \ref{eq:BtoKstar1}. In regard to the same process, looking at the numerical values of the bar diagrams of Fig.~\ref{Fig:Single-Opr-Dom-1} we find that the orders of magnitude of the BRs for LFVBDs in scenarios involving $\lQ1$, $\lq3$, and $\qe$ are quite similar and challenging to distinguish, however $\ledq$ probes a higher energy scale.

When turning our attention to the anticipated future constraints referred in the Table~\ref{Tab:LFVlimits}~(green shades in bar diagrams), 
we find that $\CRmue$, followed by $\Mueee$ to provide significant bounds on the above sets of WCs. Although there is no direct correlation between $\lQ1$, $\lq3$, $\qe$, and these LFV processes, the RGE running effects play  crucial roles in enhancing the effect of the constraints imposed by $\CRmue$ and $\Mueee$. This happens due to the influence of Higgs-lepton WCs. Consequently, the future bound for $\Mueee$ sets a cutoff on the energy scale $\Lambda$ at $\sim$ 138 TeV (as obtained from our numerical analysis) for $\lQ1$ and $\lq3$, while $\CRmue$ (Phase II) establishes a cutoff $\sim 291 \TeV$. On the other hand, if we consider the assumed future sensitivities of BRs for some of the LFVBDs (lighter  and darker red shades), we find that the dominant constraints  
come from LFVBDs
. Similarly for $\qe$, one has $\Lambda\simeq 313$~TeV from $\CRmue$ (Phase II), whereas it coincides with the scales associated with $\lQ1$ and $\lq3$ when derived from $\Mueee$.

Furthermore, investigating the expected future bounds on $\ledq$ as depicted in the bottom panel of Figure~\ref{Fig:Single-Opr-Dom-1}, we observe that the most stringent constraint emanates from $\mathcal{B}(B_s \to \mu e)$, effectively placing a cutoff on the energy scale $\Lambda$ at around 330 TeV. For other LFV decays, the future bound of $\CRmue$ closely aligns with the prospective bound of $\mathcal{B}(B_s \to \mu e)$, specifically for $[\ledq]_{1223}$. This correlation arises due to the RGE effects, that significantly contributes to $[\ledq]_{1222}$, thereby directly bolstering the constraints imposed by $\CRmue$. However, the situation differs when considering $[\ledq]_{2132}$, where there exist only some mild RGE effects.
A distinction between $[\ledq]_{1223}$ and $[\ledq]_{2132}$ emerges in that the latter contributes additionally to $\mu \to e\gamma$ and $\Mueee$. The associated RGEs of the WCs when examined reveal that the latter  contribute to the dipole operators, specifically $\cO_{eB}$ and $\cO_{eW}$, which in turn make substantial contributions to these LFV processes. It is important to note that while the current constraints from the  processes such as $\CRmue$, $\Mueee$, and $\mu \to e\gamma$ are significant, the corresponding energy scales remain below the ones established by LFVBDs. Nevertheless, if the sensitivities of the LFVBDs other than $\cB(B_s \to \mu e)$ can be enhanced by 2-4 orders of magnitude (shown in red shades),  one can probe the new physics energy scale $\sim$ $100~\TeV$ which is higher than the same probed by $\CRmue$ for both $[\ledq]_{1223}$ and $[\ledq]_{2132}$. Therefore, if new physics primarily generates the LFVBD operators above $100 \TeV$, 
we expect that $\BksZ, \BK$, $\Bsphi$, $\Bsmue$ and $\CRmue$ to be quite promising in regard to future experiments. 

We now like to enumerate the relative importance of the LFV constraints within the $\mu-\tau$ and $e-\tau$ sectors. While examining the current constraints within the said sectors (with flavor indices 2323 and 1323) as obtained in Figure \ref{Fig2:Single-Opr-Dom-1}, we find that the most stringent limitations on the WCs $\lQ1$, $\lq3$ and $\qe$ primarily arise from LFVBDs. There is hardly any necessity to perform an extended sensitivity analysis for the $\tau$ sector. The reason being, in the $\mu-\tau$ and $e-\tau$ sectors, even the most optimistic future sensitivities of BRs for all the other LFVs lie below the present limits of any of the LFVBDs. Additionally, the average reach of the excluded energy scales for these operators is much smaller compared to the same for the $e-\mu$ sector.
However, if $2q2\ell$ operators serve as the primary source of lepton flavor violation at the NP scale,  we can constrain the other LFV processes using results of this sector of LFVBDs.  In this context, as displayed in Table~\ref{Tab:correlation} we remind that the WCs $\lQ1$, $\lq3$ and $\qe$ responsible for LFVBDs contribute significantly to the Higgs-lepton WCs ($\Phil1, \phil3, \phie$) through the RGE running. We refer to Table~\ref{Table:Other LFV indirect bounds mu-tau}, for the bounds on the BRs of other LFV processes from the strongest bound available from the future limit of $\mathcal{B}(B^+ \to K^{+}\, \mu\,\tau)$ corresponding to 
the three WCs $\lQ1$, $\lq3$ and $\qe$. We find that the order of magnitude of BRs in all the processes is nearly ($\sim 10^{-12}$) which is almost 3-4 orders below the expected bounds reported by the respective experiments. 
In regard to $\Taumug$ and $\Taueg$ there are hardly any contributions from $\lQ1$ and $\qe$ compared to $\lq3$, a fact that can be attributed to the RGEs of these WCs. 
\textcolor{blue}{
 \begin{table}[h!]
 \begin{center}
 \resizebox{\textwidth}{!}{
 \renewcommand{\arraystretch}{1.3}
 \begin{tabular}{|c|c|c|c|c|c|c|}
 \hline
 Operator & BR$(Z \to \mu \tau)$ & BR$(\tau \to \mu \gamma)$ & BR$(\tau \to \mu \mu \mu)$ &  BR$(\tau \to \mu ee)$ & BR$(\tau \to \pi \mu)$ & BR$(\tau \to \rho \mu)$  \\
\hline \hline
 $[C^{(1)}_{\ell q}]_{2323}$  & $5.7\times 10^{-12}$ & $2.5\times 10^{-13}$ & $1.3\times 10^{-12}$ &  $9.7\times 10^{-13}$ & $2.6\times 10^{-12}$ & $1.4\times 10^{-12}$   \\
 \hline
 $[C^{(3)}_{\ell q}]_{2323}$  & $7.2\times 10^{-12}$ & $2.2\times 10^{-12}$ & $1.7\times 10^{-12}$ &  $1.2\times 10^{-12}$ & $3.2\times 10^{-12}$ & $1.7\times 10^{-12}$  \\
 \hline
 $[C_{qe}]_{2323}$  & $ 5.0\times 10^{-12} $  & $ 2.1\times 10^{-13} $ & $ 1.3\times 10^{-12} $ & $ 8.5\times 10^{-13}$ & $ 2.3\times 10^{-12} $ & $ 1.3\times 10^{-12} $\\
 \hline
 \end{tabular}}
 \caption{Indirect upper limits on the 
 $\BR$s of the other LFV processes in the $\mu-\tau$ sector, obtained from the future limit of $\mathcal{B}(B^+ \to K^{+}\, \mu\,\tau)$ (strongest constraint) in the 1-D scenario. 
 }
 \label{Table:Other LFV indirect bounds mu-tau}
 \end{center}
 \end{table}
}

Coming back to the $e-\mu$ sector we now like to explore the prospect of constraining LFVBDs 
via the relevant WCs $\lQ1$, $\lq3$ and $\qe$ and $\ledq$ while considering 
the future limits of $\Mueee$, and $\CRmue$ (Phase-I and II). The requirement for finding these numbers in Table~\ref{Table:LFVBD indirect bound} is that, the BRs of these processes which are also induced by the WCs' listed in column 2 of the same table should lie below their future limits. Therefore focusing on these numbers we observe that 
from the future bounds of $\CRmue$ (Phase I), in scenarios like $\lQ1$, $\lq3$ and $\qe$, the upper limits of $\mathcal{B}(B \to K^{(*)} \mu e)$ and $\mathcal{B}(B_s \to \phi \mu e)$ ($\sim 10^{-10}$) are smaller by one order of magnitude compared to the current experimental bounds. It is interesting to note that these numbers fall within the scope of the future bounds that are expected to be obtained through upgrades at LHCb and Belle II as discussed in Section~\ref{Sec:Intro}. This also falls within a 2-order smaller LFVBDs zone as explored in this work. However, the situation is notably different for $\mathcal{B}(B_s \to \mu e)$, since the above-mentioned WCs predict too low BRs that are way beyond the reach of B factories. This is in contrast to the case of $\ledq$ scenarios. Here, considering limits as obtained from $\CRmue$ (Phase I), the BR bounds LFVBDs are significantly higher than the current limits, and no data are shown in this regard. For the process $\Mueee$ in regard to $\lQ1$, $\lq3$ and $\qe$ we also observe a similar 
pattern of a lowered BR. The bounds obtained from $\Mueee$ are about one order of magnitude stronger than those from $\CRmue$ (Phase I), allowing for even tighter constraints on the BRs of LFVBDs. We point out that the $\ledq$ scenarios do not affect $\Mueee$ through RGEs, thus there is no data corresponding to its entries in the $\Mueee$ column of Table~\ref{Table:LFVBD indirect bound}.

Moving forward, the predicted future bound of $\CRmue$ (Phase II) provides the most stringent constraints on these WCs compared to the other two scenarios. Consequently, for $\lQ1$, $\lq3$ and $\qe$, the indirect UL on the BRs of LFVBDs is further reduced by several orders of magnitude for $\mathcal{B}(B \to K^{(*)} \mu e)$,  $\mathcal{B}(B_s \to \phi \mu e)$ and $\mathcal{B}(B_s \to \mu e)$ compared to the current limits. 
However, in the case of $\ledq$, a slightly different pattern emerges. The indirect ULs obtained in this scenario for $\mathcal{B}(B \to K^{(*)} \mu e)$ and $\mathcal{B}(B_s \to \phi \mu e)$ decays are $\sim 10^{-12}$, while for $\mathcal{B}(B_s \to \mu e)$ the UL is $\sim 10^{-10}$. The later UL 
aligns with the anticipated future limit proposed by LHCb-II (Table~\ref{Tab:LFVlimits}).

\begin{table}[h!]
\begin{center}
\resizebox{\textwidth}{!}{
\renewcommand{\arraystretch}{1.3}
\begin{tabular}{|c|c|c|c|c|}
\hline
Observable & WC & UL from BR$(\mu \to eee)$ & UL from CR$(\mu \to e,\,\text{Al})$, Phase I & UL from CR$(\mu \to e,\,\text{Al})$, Phase II \\
\hline \hline
\multirow{4}{*}{BR$(B^+ \to K^+ \mu^-e^+)$}
& $[C^{(1)}_{\ell q}]_{1223}$ & $2.9\times 10^{-11} $  & $2.2\times 10^{-10}$ &  $1.5\times 10^{-12} $   \\
& $[C^{(3)}_{\ell q}]_{1223}$ & $1.9\times 10^{-11} $   & $1.5\times 10^{-10}$ &  $9.8\times 10^{-13} $   \\
& $[C_{qe}]_{2312}$           & $ 3.2\times 10^{-11} $  & $1.8\times 10^{-10}$ &  $ 1.2\times 10^{-12} $ \\
& $[C_{ledq}]_{1223}$         & -                         & -                      &  $1.9\times 10^{-11}$ \\
\hline
\multirow{4}{*}{BR$(B^0 \to K^{*0}\mu^-e^+)$}
& $[C^{(1)}_{\ell q}]_{1223}$  & $6.3\times 10^{-11} $  & $4.7\times 10^{-10}$ &  $3.4\times 10^{-12} $   \\
& $[C^{(3)}_{\ell q}]_{1223}$  & $4.2\times 10^{-11} $  & $3.3\times 10^{-10}$ &  $2.3\times 10^{-12} $    \\
& $[C_{qe}]_{2312}$            & $ 6.9\times 10^{-11} $ & $3.8\times 10^{-10}$ &  $ 2.9\times 10^{-12} $ \\
& $[C_{ledq}]_{1223}$          & -                        & -                      &  $7.9\times 10^{-12}$ \\
\hline
\multirow{4}{*}{BR$(B_s\to \phi \mu^-e^+)$}
& $[C^{(1)}_{\ell q}]_{1223}$  & $6.7\times 10^{-11} $  & $4.9\times 10^{-10}$ & $3.4\times 10^{-12} $   \\
& $[C^{(3)}_{\ell q}]_{1223}$  & $4.5\times 10^{-11} $  & $3.5\times 10^{-10}$ &  $2.3\times 10^{-12} $   \\
& $[C_{qe}]_{2312}$            & $ 7.5\times 10^{-11} $ & $4.1\times 10^{-10}$ &  $2.8\times 10^{-12} $ \\
& $[C_{ledq}]_{1223}$          & -                        & -                      &  $8.6\times 10^{-12}$ \\
\hline
\multirow{4}{*}{BR$(B_s\to \mu^-e^+)$} 
& $[C^{(1)}_{\ell q}]_{1223}$ & $8.0\times 10^{-14}$   & $5.9\times 10^{-13}$ &  $4.0\times 10^{-15}$  \\
& $[C^{(3)}_{\ell q}]_{1223}$ & $5.3\times 10^{-14} $  & $4.2\times 10^{-13}$ &  $2.7\times 10^{-15} $   \\
& $[C_{qe}]_{2312}$           & $ 8.9\times 10^{-14} $ & $4.9\times 10^{-13}$ &  $3.4\times 10^{-15} $   \\
& $[C_{ledq}]_{1223}$         & -                        & -                      &  $1.4\times 10^{-10} $   \\
\hline
\end{tabular}}
\caption{Indirect upper limits from $\mu \to eee$, CR$(\mu \to e,\,\text{Al})$ Phase I and II on different LFVBD processes considering a single operator responsible for such processes, at the scale $\mu=\Lambda$. See text for the absence of data for $\ledq$.}\label{Table:LFVBD indirect bound}
\end{center}
\end{table}


\subsection{Two Operators Interference}

In this section we discuss the effects when two operators are turned on at the high energy scale $\Lambda$, particularly in relation to how their mutual interference can affect the different LFV observables including LFVBDs. Prime motivations for such analysis are twofold. First, we can analyze the operator mixing and interplay of RGE flows among the operators which are responsible exclusively for LFVBDs. Second, we can estimate the effects of the non-trivial cancellations due to appropriately chosen different pairs of WCs that appear in the BR formulae of LFV processes of interest. Since  WCs are related to each other via RGEs, they can combine in different strengths to suppress or enhance concerned LFV processes. 

In the plane of a pair of WCs, the so-called \textit{flat-directions} are defined as the portions of the contour where there is a cancellation from the contributing terms within the BR arising from these WCs~\cite{Calibbi:2021pyh}. In such scenarios it is also possible to tune the parameters to enhance the relative strengths of LFV processes. Such relations can easily be found from the BR formula of the corresponding LFV process and they would appear as cuspy regions in the 2-dimensional plots in this analysis. For our purpose, from equations \ref{eq:Bs to l1l2}-\ref{eq:B to Kstar-l1l2}, corresponding forms of WCs given in equations \ref{eq:C9}-\ref{eq:CSP} and considering only the relevant operators that contribute to LFVBDs, we find that

\bea
\Br\left[ B_s \to \ell^+_i \ell^-_j \right] &\sim& 
k_1 \bigg\{k_2 \bigg(\lQ1+\lq3+\qe\bigg) + k_3 \bigg(\ledq-\ledqp \bigg)\bigg\}^2
\nonumber\\ &+& k_4 \bigg\{k_5 \bigg(\lQ1+\lq3-\qe \bigg) + k_6 \bigg(\ledq+\ledqp \bigg)\bigg\}^2\,\label{eq:Bstomue1},\\ 
\Br \left[B^0\to K^{(*)}\ell_i^+\ell_j^-\right] &\sim& 
k_7 \bigg\{\bigg(\lQ1+\lq3\bigg)^2 +\bigg(-\qe\bigg)^2\bigg\} \nonumber\\ 
&+& k_8 \bigg\{\bigg(\lQ1+\lq3\bigg)^2 + \bigg(\qe\bigg)^2\bigg\} \nonumber\\ 
&+& k_9 \bigg({\ledq}-{\ledqp}\bigg)^2,
\label{eq:BtoKstar1}
\eea 
where $k_i$s refer to appropriate products from 
Eq.~\ref{eq:Bs to l1l2}. Considering 2D cases the above equations show that for both types of LFVBD processes, one may have interfering terms containing a pair of WCs. 

In the following discussion, depending on the possibility of having flat-directions in the BRs we analyze below five different 2D scenarios. Among them, we study the cancellation within the pair of WCs responsible for LFVBDs in the first three scenarios. For this, we simultaneously consider the presence of two non-zero coefficients of $2q2\ell$ operators relevant for LFVBDs at the scale $\Lambda$. Considering the operators with significant effects to LFVBDs and using Eq.~\ref{eq:Bstomue1} and~\ref{eq:BtoKstar1} as our guiding principle, we choose the scenarios with $[C_{\ell q}^{(1)}]_{1223}$ - $[C_{\ell q}^{(3)}]_{1223}$, $[C_{\ell edq}]_{2132}$ - $[C_{\ell q}^{(1)}]_{1223}$ and $[C_{\ell edq}]_{1223}$ - $[C_{qe}]_{2312}$.
Similarly, to study the possibility of non-trivial cancellations, we simultaneously consider i) the presence of a non-zero coefficient of $2q2\ell$ operators relevant for LFVBDs (either one from $[C_{\ell q}^{(1,3)}]_{1223}$ 
,$[C_{qe}]_{2312}$, $[C_{\ell edq}]_{1223}$) and ii) non-vanishing Higgs-lepton operators  
that are relevant for $\mu \to eee$, CR$(\mu \to e,\,\text{Al})$.

\begin{figure}[ht!] 
\centering
\includegraphics[width=12cm,height=7cm]{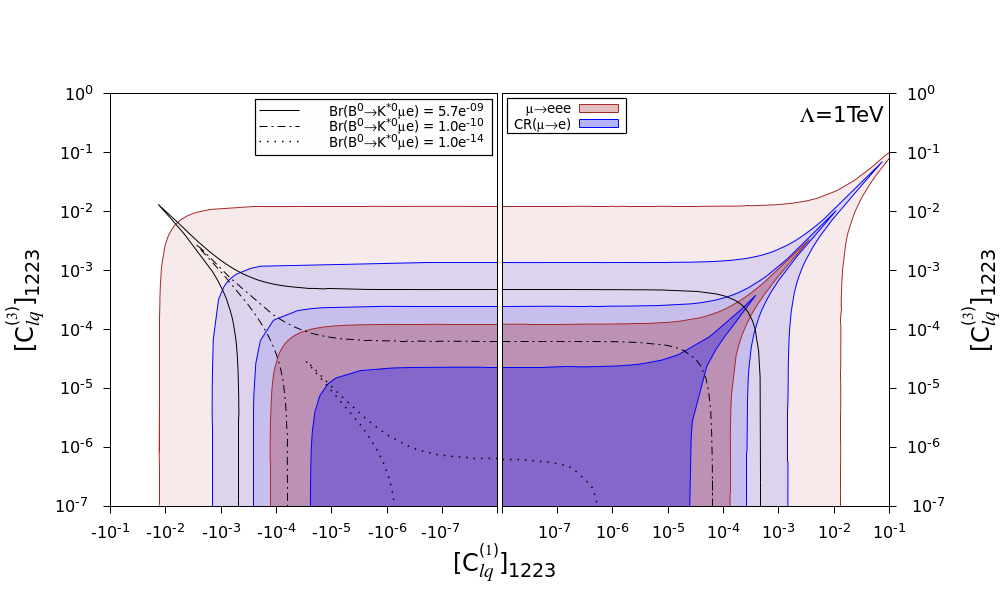}
\caption{With $\Lambda=1$~TeV, contours for $B^0 \to K^{*0}\mu\e$ as a function of $[C_{\ell q}^{(1)}]_{1223}$ and $[C_{\ell q}^{(3)}]_{1223}$. Lighter colors show the currently allowed regions for $\Mueee$ (brown) and $\CRmue$ (blue) bounds whereas darker shades represent futuristic limits. For $\CRmue$ the lighter and darker blue shades correspond to Phase I ($\BR \sim 10^{-15}$) and Phase II ($\BR \sim 10^{-17})$ respectively.}
\label{Plot:clq1_clq3}
\end{figure}

     Figure~\ref{Plot:clq1_clq3} illustrates the contours of  $\cB(B^0 \to K^{*0} \mu e), \Mueee$, and $\CRmue$ in the plane of WCs $[\lQ1]_{1223}$ and $[C_{\ell q}^{(3)}]_{1223}$, both referring to the energy scale $\Lambda =1 \TeV$.  It is observed that similar results hold true also for higher values of $\Lambda$, with only minor logarithmic modifications due to RGEs. Lighter colors show the currently allowed regions for $\Mueee$ (brown) and $\CRmue$ (blue) experiments whereas darker shades of the same colors represent futuristic experimental limits tabulated in Table \ref{Tab:LFVlimits}. For $\CRmue$, we differentiate between two future expectations: the slightly lighter blue shade corresponds to Phase I ($\BR \sim 10^{-15}$) as proposed by J-PARK and Fermi Lab \cite{HernandezVillanueva:2022dpt} and the darker blue shade corresponds to Phase II ($\BR \sim 10^{-17}$ \cite{HernandezVillanueva:2022dpt}). The contours for $B^0 \to K^{*0}\mu e$ are shown in black, with the solid line indicating the current limit ($\BR \sim 10^{-9}$), which is also the most stringent limit among these three processes for this set of WCs. However, the future predictions for $\Mueee$ and $\CRmue$ (Phase I) impose stronger constraints on both $[\lQ1]_{1223}$ and $[C_{\ell q}^{(3)}]_{1223}$. Notably, $\Mueee$ surpasses Phase I of $\CRmue$ in terms of constraining these WCs. The new parameter space constrained by $\Mueee$ for this set of WCs reduces $\mathcal{B}(B^0 \to K^{*0} \mu e)$ by one order of magnitude from the existing bound. Belle II and LHCb experiments can potentially probe this limit. Moreover, $\rm CR(\mu \to e)$ (Phase II) impose strongest constraints on the parameter space of $[\lQ1]_{1223}$ and $[C_{\ell q}^{(3)}]_{1223}$ causing $\mathcal{B}(B^0 \to K^{*0}\, \mu e)$ to shrink 2-3 order of magnitude from the existing bound. However, achieving this level of BR is very challenging given the current and the upcoming scopes of B factories in the near future. The plot in the $[\lQ1]_{1223}$ and $[C_{\ell q}^{(3)}]_{1223}$ plane also provides insights into the flat directions among all three LFV processes. The flat direction for $B^0 \to K^{0*} \mu e$ fall in the second quadrant and the associated  cancellation is consistent with Eq.~\ref{eq:BtoKstar1}. In contrast, for the other LFV processes namely,  $\Mueee$ and $\CRmue$ cancellations occur on the opposite side. We note that the 
     processes like $\Mueee$ and $\CRmue$ exhibit flat directions resulting from cancellations between $[C_{\phi \ell}^{(1)}]_{12}$ and $[C_{\phi \ell}^{(3)}]_{12}$ and this is induced by the RGE flows.
     The latter can be understood by analysing the running and matching of SMEFT and LEFT operators. 
     Using appropriate results from Sec.~\ref{Sec:Low-energy Eff. Hamiltonian} we find,
      \be
     \label{Eq:WC RGEs}
     \big[C_{\phi \ell}^{(1)}(\mu) + C_{\phi \ell}^{(3)}(\mu)\big]_{12} \approx \frac{3Y_c Y_t}{8\pi^2}\mathrm {log}(\frac{\mu}{\Lambda}) \big[C_{\ell q}^{(1)}(\Lambda)-C_{\ell q}^{(3)}(\Lambda)\big]_{1223}.
     \ee
     The presence of heavy Yukawa terms with top quark in Eq.~\ref{Eq:WC RGEs} represents the fact that large Higgs-lepton operators can be induced by the $2q2\ell$ operators, even though they were absent at the starting scale $\Lambda$.  
     Additionally, in regard to the contours corresponding to $\Mueee$ and $\CRmue$ cuspy regions arise in the first quadrant of the figure drawn in $\lQ1$-$\lq3$ plane. Thus, a small value of the right hand side of Eq.~\ref{Eq:WC RGEs}
     for the said  region 
     corresponds to 
     a small value for the sum $
     C_{\phi \ell}^{(1)}(\mu) + C_{\phi \ell}^{(3)}(\mu)$
    . It may easily be seen that the same  would 
     appear in the BR formulae of $\Mueee$ and $\CRmue$, 
    that in turn means a flat direction or cancellation of appropriate terms induced by RGE effects. A close examination of the plot also reveals that the cuspy region resulting from the current limit of $B^0 \to K^{*0} \mu e$ mildly constrains the parameter space allowed by future bounds of $\Mueee$ and $\CRmue$ (Phase I). Similarly, the cuspy region stemming from the current bound of $\CRmue$ limits some of the regions allowed by the $B \to K^{*0} \mu e$ contour. Likewise, the converse is true in the left quadrant.
    
 \begin{figure}[ht!] 
\centering
\includegraphics[width=12cm,height=7cm]{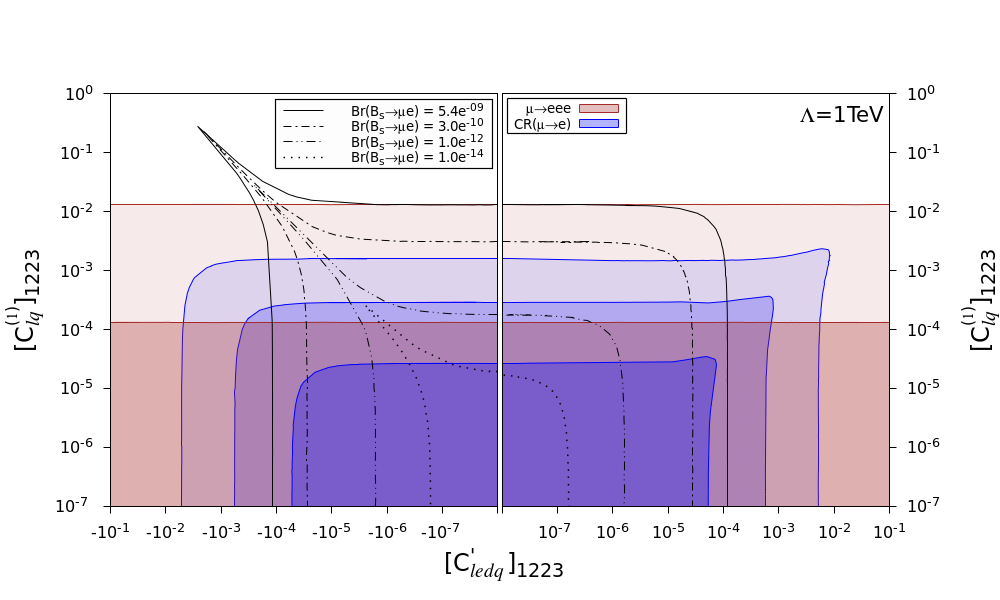}
\includegraphics[width=12cm,height=7cm]{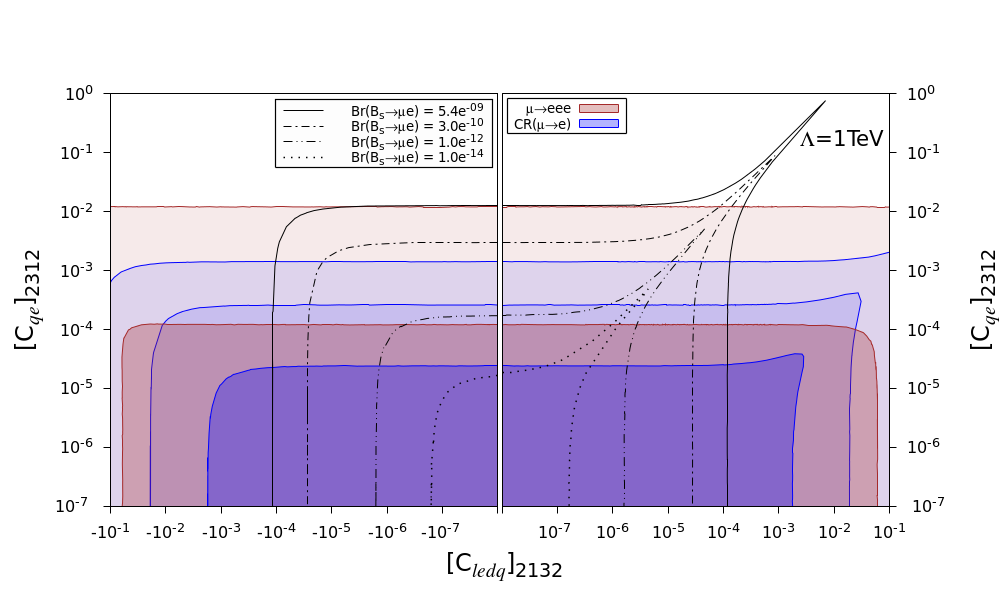}
\caption{With $\Lambda=1$~TeV, contours of $B_s \to \mu\,e$ as a function of $[C_{\ell edq}]_{1223}$ and $[C_{\ell q}^{(1)}]_{1223}$. Color codes are same as in Fig.~\ref{Plot:clq1_clq3}.}
\label{Plot:cledq_clq1}
\end{figure}  

   Fig.~\ref{Plot:cledq_clq1} represents the contours of  $\cB(B_s \to \mu\,e), \Mueee$ and $\CRmue$, in the plane of WCs $[C'_{\ell edq}]_{1223}$ vs $[C_{\ell q}^{(1)}]_{1223}$ (Top) and $[C_{\ell edq}]_{2132}$ vs $[C_{qe}]_{2312}$ (Bottom), both corresponding to the energy scale $\Lambda =1 \TeV$. All the colored contours have similar classifications as described in the above paragraph. One noteworthy aspect of these WCs is that, at the low-energy limit $m_b$, they simultaneously influence $C_9$ and $C_{10}$, as well as $C_S^{(\prime)}$ and $C_P^{(\prime)}$. In comparison to Figure~\ref{Plot:clq1_clq3}, we focus here on the contours corresponding to $B_s \to \mu\,e$ rather than $B \to K^{(*)}\, \mu e$. This choice is justified by the fact that $B_s \to \mu\,e$ places stronger constraints on $[C'_{\ell edq}]_{1223}$ and $[C_{\ell edq}]_{2132}$, as clearly evident from Fig.~\ref{Fig2:Single-Opr-Dom-1}. Analyzing the contours associated with the current limits for all the three processes, we observe that the strongest constraint along the x-axes comes from $B_s \to \mu\,e$, however, along the y-axes it comes from $\CRmue$. Turning to future limits, the current limit of $B_s \to \mu\,e$ surpasses $\CRmue$ (Phase I) along the x-axes, while along the y-axes, both $\CRmue$ (Phase I) and $\Mueee$ exert strong constraints. The limit from $\CRmue$ (Phase II), while being the most influential constraint in the top plot for both WCs, slightly lags behind the current limit of $B_s \to \mu\,e$ along the $[C_{\ell edq}]_{2132}$ direction in the bottom plot. We notice that, the contours of $\Mueee$ have negligible effects along $[C'_{\ell edq}]_{1223}$, since the RGE effects of $[C'_{\ell edq}]_{1223}$ has a minimal impact on the WCs contributing to $\Mueee$. However, it does contribute to  $\CRmue$ through $[C_{\ell edq}]_{1222}$, as shown in Table~\ref{Tab:correlation}.  On the other hand, in the bottom plot, $[C_{\ell edq}]_{2132}$ mildly affects both $\Mueee$ and $\CRmue$ through dipole operators. Consequently, the current and future constraints on $\mu \to eee$ are notably weaker along the x-axes when it comes in constraining both $[C'_{\ell edq}]_{1223}$ and $[C_{\ell edq}]_{2132}$. This trivial cancellation observed between the pair of WCs, in both the top and bottom figures, can be attributed to the cross-terms appearing in Eq. \eqref{eq:Bs to l1l2}. However, prominent cancellations are not observed in the other LFV processes, primarily due to the absence of cross-terms between the corresponding WCs responsible for both $\CRmue$ and $\Mueee$ that are affected by the RGE flow. 

\begin{figure}[ht!] 
\centering
\includegraphics[width=12cm,height=7cm]{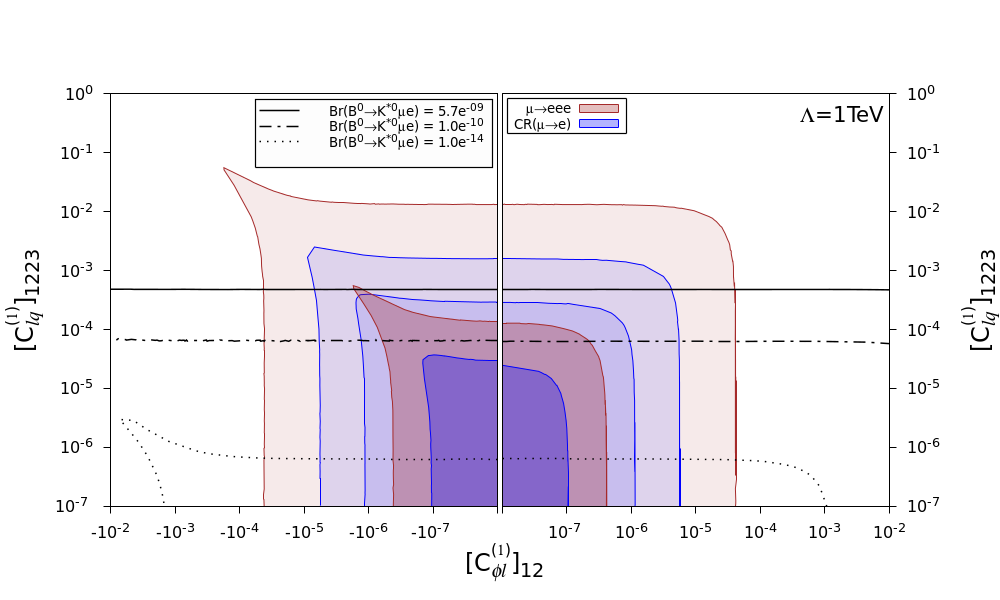}
\includegraphics[width=12cm,height=7cm]{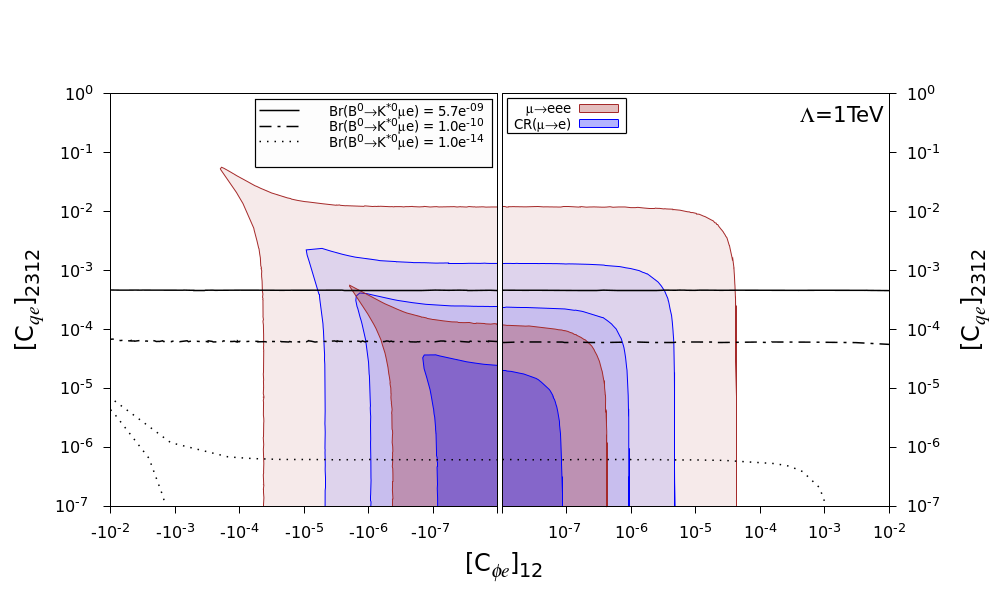}
\caption{With $\Lambda=1$~TeV, contours of $B^0 \to K^{0*}\, \mu\,e$ as a function of $[C_{\phi \ell}^{(1)}]_{12}$ and $[C_{\ell q}^{(1)}]_{1223}$ (top panel) and $[C_{\phi e}]_{12}$ and $[C_{qe}]_{2312}$ (bottom panel). Color codes are same as in Fig.~\ref{Plot:clq1_clq3}.}
\label{Plot:cphi_clq1cqe}
\end{figure}

Figure~\ref{Plot:cphi_clq1cqe} displays the contours of $\cB(B^0 \to K^{*0}\, \mu e)$, $\Mueee$ and $\CRmue$, in the plane of WCs $[C_{\phi \ell}^{(1)}]_{12}$ and $[\lQ1]_{1223}$ (top) and $[C_{\phi e}]_{12}$ and $[C_{qe}]_{2312}$ (bottom), both considered at the energy scale $\Lambda =1 \TeV$. In this context, $[\lQ1]_{1223}$ and $[C_{qe}]_{2312}$ pertain to LFVBDs, while $[C_{\phi \ell}^{(1)}]_{12}$ and $[C_{\phi e}]_{12}$ are relevant for $\Mueee$ and $\CRmue$. The color codes for the contours hold the same meaning as described in the previous paragraph. Looking into the current limits of the BRs, we find that the $\CRmue$ process imposes the strongest constraint along $[C_{\phi \ell}^{(1)}]_{12}$ (top) and $[C_{\phi e}]_{12}$ (bottom). However, along the y-axes, the current limit of $\cB(B^0 \to K^{*0} \mu e)$ surpasses the other LFV processes in constraining $[\lQ1]_{1223}$ (top) and $[C_{qe}]_{2312}$ (bottom) respectively. Furthermore, considering future bounds we find that the contours of $\Mueee$ and $\rm CR(\mu \to e)$ (Phase I) strongly constrain the WCs along x as well as y axes in both the plots. These future bounds from both the LFV processes closely coincide with the current bound of $\cB(B^0 \to K^{*0}\, \mu e)$, especially in constraining $[\lQ1]_{1223}$ (top) and $[C_{qe}]_{2312}$ (bottom). Moreover, the future limit of $\CRmue$ (Phase II), although imposes strong constraints on $[C_{\phi \ell}^{(1)}]_{12}$ (top) and $[C_{\phi e}]_{12}$ (bottom), is less efficient in constraining both $[\lQ1]_{1223}$ (top) and $[C_{qe}]_{2312}$ (bottom). We note that in regard to $\CRmue$ and $\Mueee$, the observed pattern of cancellations between pairs of WCs in the top figure can again be understood from Eq.~\ref{Eq:WC RGEs}. The large contributions to both Higgs-lepton operators in Fig.~\ref{Plot:cphi_clq1cqe} coming from $2q2\ell$ operators completely explains why $\mu \to eee$ and $\CRmue$ processes dominates in constraining these operators. On the contrary, for the $2q2\ell$ operators, strong constraints mainly arise from LFVBDs. In regard to the bottom figure, a similar equation relating $[\phie]_{12}$ and $[\qe]_{2312}$ can be found from corresponding RGEs to explain their behaviour seen in the plot.  
    

\section{Conclusion}
\label{Sec:Conclusion}

  Using a model-independent framework of SMEFT we analyzed how the limits from leptonic and semi-leptonic LFV B-decay (referred to as LFVBD) processes such as $\Bsmue, \BK, \BksZ, \Bsphi$ 
  may constrain relevant Wilson Coefficients associated with dimension-6 operators as mentioned in Table~\ref{tab:BtoS operators1}. We use the usual match and run procedure where 
  a SMEFT journey of appropriate Wilson coefficients from a high scale ($\Lambda$) to the electroweak scale would determine the starting of the evolution for 
  the LEFT operators.  The latter are then evolved down to a low scale like the mass scale $m_b$ used to obtain the appropriate branching ratios. 
  Apart from the said LFVBD limits, we also take into account how the bounds from LFV processes like   
  $\CRmue$, $\ell_i \to \ell_j \gamma$, $\ell_i \to\ell_j\ell_k\ell_m$ and $Z \to \ell_i \ell_j$, collectively referred as the other LFV processes,  are able to constrain
  the WCs. 
  We used present BRs as well as some expected future limits as given in Table-\ref{Tab:LFVlimits} while probing the WCs of importance and studied the
  interplay of the LFVBD and the other LFV processes. While direct effects from LFVBDs principally affect the associated WCs, 
  the same set of the coefficients may receive constraints from the other LFV processes, too. 
  Accordingly, the running effects and RGE mixing of different WCs can be quite important while we do a combined analysis of all these
  LFV processes. For each of the relevant SMEFT operators we identify a few WCs that are heavily influenced because of RGE
  mixing and running and are in turn able to affect the different LFV processes considered in this analysis.  
  In addition to studying the direct effects of LFVBD limits on 
  the primarily important WCs, we also take into account how the same are
  affected indirectly via 
  BR limits from the other LFV processes. The latter set of LFV limits, on the
  other hand, may provide estimates for the maximum levels of the LFVBDs that
  may duly be probed. Additionally, apart from considering prospective limits from Table-\ref{Tab:LFVlimits} we also considered two assumed levels of
  more stringent LFVBD limits simply to explore the potential of the interplay
  of all the LFV processes under discussion.  
  We divide our analysis into two principal parts, namely, studying
  only one or two operators at a time. 
  In the first part, the single operator analysis identifies the WCs that are maximally affected 
  via RGE mixing and running effects for a given 
  SMEFT scale ($\Lambda=1$~TeV). We further estimated maximum possible values of $\Lambda$  
  as obtained from a given LFV limit while assuming the concerned WC to be unity.

 In regard to the RGE effects of LFVBD operators it turns out that only $\cO_{\vp\ell}^{(1,3)}$ and $\cO_{\vp e}$, which affect most of the other LFV processes, receive significant impacts from $\cO_{\ell q}^{(1,3)}$ and $\cO_{qe}$. This correlation is shown in the Table~\ref{Tab:correlation}.  For $\cO_{ledq}$, the prominent RGE effect of $[C_{ledq}]_{1223}$ arises from the same WC with different quark index, namely $[C_{ledq}]_{1222}$. Out of the six operators primarily associated with LFVBDs as shown in Table~\ref{tab:BtoS operators1}, one finds from the 1-D analysis 
(performed with $\Lambda=1$~TeV) that only the operators with left-handed quark currents, such as $C_{\ell q}^{(1,3)}$ and $\qe$ are able to contribute to the other LFV processes
insignificantly (Fig.~\ref{Fig:Single-Opr-Dom-1}). In contrast, the operators with right-handed quark currents namely 
$\ld$ and $\ed$ that contribute to $C_{(9,10)}^{\prime}$, are relevant only for LFVBDs, with hardly having any effect
on the other LFV processes.

The later part of the 1-D analysis for the LFV operators that probed the scale $\Lambda$ while considering the current and future experimental bounds results into Fig.~\ref{Fig:Single-Opr-Dom-1}. Major take away from this energy analysis is that, in the $e-\mu$ sector, current sensitivities (blue bars) of BRs of LFVBDs coming from operators $\cO_{\ell q}^{(1,3)}, \cO_{qe}$ and $\cO_{ledq}$, are competitive  compared to other similar low-energy observables, specially with $\rm CR (\mu\rightarrow e)$ or sometines with $\Mueg$. The result remains the same if we consider future (green shades) or assumed enhancement of sensitivities (darker and lighter red shades). Therefore, if new physics primarily generates the LFVBD operators between the scales $\Lambda=100 - 1000$~$\TeV$, we expect that $\BksZ, \BK$, $\Bsphi$, $\Bsmue$ and $\CRmue$ to be quite promising in regard to future experiments. 

In this sector, two processes, namely $\Mueee$ and $\CRmue$ can put indirect constraints on $\BR$s of several $B$-decay processes as shown in the Table~\ref{Table:LFVBD indirect bound}. From this analysis we find that for the processes $\BK, \BKs$ and $\Bsphi$, the $\BR$s $\sim 10^{-10}$ which is just one order of magnitude below the current LHCb bounds and within the anticipated future limit proposed by LHCb-II. On the other hand, the $\ell-\tau$ sector is more promising as they can be probed at much lower energy, clearly seen from the plots of Fig.~\ref{Fig2:Single-Opr-Dom-1}.

 In the 2-D case, we consider two non-vanishing operators at a time for a fixed value of $\Lambda=1$~TeV.  Both WCs may
  directly be related to LFVBDs (Figs.~\ref{Plot:clq1_clq3},~\ref{Plot:cledq_clq1}) or one related to LFVBD and another corresponding to
  a different LFV process other than any of the LFVBDs (Fig.~\ref{Plot:cphi_clq1cqe}).
  We begin with considering a pair of WCs responsible for LFVBDs only. Fig.~\ref{Plot:clq1_clq3} shows a plot for 
  the WCs in the plane of $\lQ1$ and $\lq3$ where present and future BRs of LFV processes like $\BKs$, $\CRmue$ and $\Mueee$ are used for the contours.
Following the current limits we find that, although $\BKs$ imposes the strongest constraints on these two WCs, future predictions for $\Mueee$ and Phase I of $\CRmue$ overcome these limits. To be specific, the new parameter space constrained by $\Mueee$ for this set of WCs reduces the $\mathcal{B}(B \to K^* \mu e)$ by one order of magnitude from the existing bound that can potentially be probed by Belle II and LHCb experiments. Similarly, in Fig.~\ref{Plot:cledq_clq1} we show the contours for $\Bsmue$, $\CRmue$ and $\Mueee$ in the plane of
  $[C_{\ell edq}]_{1223}$ vs $[C_{\ell q}^{(1)}]_{1223}$ and $[C_{\ell edq}]_{2132}$ vs $[C_{qe}]_{2312}$.  The figures show that
  RGE running of $[C_{\ell edq}]_{1223}$ has very little impact on the WCs contributing to $\Mueee$ whereas its influence on $\CRmue$ via
  $[C_{\ell edq}]_{1222}$ is relatively more significant. This is consistent with the result shown in Table~\ref{Tab:correlation}.
  Moreover, it turns out that $[C_{\ell edq}]_{2132}$ has mild influence on both $\Mueee$ and $\CRmue$ through dipole operators.

In further study of our 2-D analysis we picked up a pair of operators, of which one is exclusively responsible for LFVBDs and the other one is significant for several other LFV processes. From Fig.~\ref{Plot:cphi_clq1cqe} we find that, while constraining $[C_{\ell q}^{(1)}]_{1223}$ and $[C_{qe}]_{2312}$, future bounds from $\Mueee$ and $\rm CR(\mu \to e)$ (Phase I) closely coincide with the current bound of $\cB(B^0 \to K^{*0}\, \mu e)$. This imply that different LFVBDs processes are competitive in order to impose constraints on the $2q2\ell$ operators. On the other hand, Higgs-lepton operators get strongest constraints from other LFV processes only. The flat-directions in these plots are indicative of the non-trivial RGE effects between the respective pair of operators.

\appendix

\section{ Lepton flavor violating $Z$ boson decays  ($\boldsymbol{Z\rightarrow \ell_i\ell_j}$)}
\label{Appendix: ZLFV decays}

The effective interactions involving the $Z$ boson and the SM leptons, including those responsible for LFV effects, are given by the following Lagrangian~\cite{Brignole:2004ah}
\begin{align}
\mathcal L_{\rm eff}^{Z} = &
\Big[ \left(g_{VR}\, \delta_{ij} +\delta g^{ij}_{VR}\right)~ \bar\ell_i\gamma^\mu P_R\ell_j 
\,+\, \left(g_{VL}\, \delta_{ij} +\delta g^{ij}_{VL}\right) ~ \bar\ell_i\gamma^\mu P_L \ell_j \Big] Z_\mu ~+ \nonumber \\
& \Big[\delta g^{ij}_{TR}~ \bar\ell_i\sigma^{\mu\nu} P_R\ell_j\, + \,g^{ij}_{TL}~ \bar\ell_i\sigma^{\mu\nu} P_L\ell_j \Big]  Z_{\mu\nu} ~ + ~ h.c. \,,
\end{align}
where 
\be
g_{VR} =\frac{e\sw}{\cw}
\,,\quad\quad g_{VL} = \frac{e}{\sw\cw} \left(-\frac{1}{2} + \sw^2\right)\,,
\ee
are the SM couplings of the $Z$ to, respectively, right-handed (RH) and left-handed (LH) lepton currents, with $\sw$ ($\cw$) being the sine (cosine) of the weak mixing angle.
New physics effects are encoded in the effective couplings $\delta g_{V/T}$, which at the tree level match the SMEFT operators as follows
\begin{align}
\label{eq:gV}
  &\delta g^{ij}_{VR} = - \frac{ev^2}{2\sw\cw \Lambda^2} \, C^{ij}_{\vp e} \,,
   \quad
    \delta g^{ij}_{VL} = - \frac{ev^2}{2\sw\cw \Lambda^2} \, 
   \Big(C^{(1)\, ij}_{\vp\ell}+C^{(3)\, ij}_{\vp\ell}\Big)\,,
   \\
&    \delta g^{ij}_{TR} = \delta g^{ji\,*}_{TL} = -\frac{v}{\sqrt2 \Lambda^2}\,
    \Big(\sw C_{eB}^{ij} + \cw C_{eW}^{ij}\Big)\,,
    \label{eq:gT}
\end{align}
where the WCs have to be evaluated at the scale $\mu=m_Z$.

The branching ratios of the $Z$ decays into leptons, in particular of the LFV modes, are then given by the following expression~\cite{Brignole:2004ah,Crivellin:2013hpa}
\begin{align}
\mathrm{BR}\left( {Z \to \ell_i \ell_j } \right) =
\frac{m_Z}{12\pi \Gamma_Z} \Bigg(&
 \left| g_{VR}\, \delta_{ij} + \delta g^{ij}_{VR} \right|^2 + \left| g_{VL}\, \delta_{ij} + \delta g^{ij}_{VL} \right|^2  \nonumber \\
 &+ \frac{m_Z^2}{2} \left| \delta g^{ij}_{TR} \right|^2 +
  \frac{m_Z^2}{2} \left| \delta g^{ij}_{TL} \right|^2
  \Bigg)\,,
  \label{eq:Zll}
\end{align}
where $\Gamma_Z = 2.4952(23)$~GeV is the total decay width of the $Z$~boson, and we summed over the two possible combinations of lepton charges, $\ell^\pm_i \ell^\mp_j$.

\section*{Acknowledgements}
We have been benefited from discussion with X. Marcano. IA would like to thank the Council of Scientific and Industrial Research, India for financial support.
\printbibliography

\end{document}